\documentclass[%
aip,pop,amsmath,amssymb,
 reprint,%
]{revtex4-1}

\pdfoutput=1
\usepackage{color}
\usepackage{xcolor}
\usepackage{graphicx}
\usepackage{bm}
\newcommand{\subfigimg}[3][,]{%
  \setbox1=\hbox{\includegraphics[#1]{#3}}
  \leavevmode\rlap{\usebox1}
  \rlap{\hspace*{200pt}\raisebox{\dimexpr\ht1-2\baselineskip}{#2}}
  \phantom{\usebox1}
}

\begin{document}

\preprint{AIP/123-QED}

\title[]{On the compressibility effect in test particle acceleration 
by magnetohydrodynamic turbulence}
\author{C.A. Gonz\'alez}
 \email{caangonzalez@df.uba.ar}

\author{P. Dmitruk}
\author{P.D. Mininni}
 \
\affiliation{Departamento de F\'isica, Facultad de Ciencias Exactas y Naturales, Universidad
de Buenos Aires and IFIBA, CONICET, Ciudad universitaria, 1428 Buenos Aires, Argentina}
\author{W.H. Matthaeus}
\
\affiliation{Bartol Research Institute and Department of Physics and Astronomy, University of
Delaware, Newark, Delaware, USA}


\begin{abstract}
The effect of compressibility in charged particle energization by 
magnetohydrodynamic (MHD) fields is studied in the context of test
particle simulations. This problem is relevant to the solar wind and
the solar corona due to the compressible nature of the flow in those
astrophysical scenarios. We consider turbulent electromagnetic fields
obtained from direct numerical simulations 
of the MHD equations with a strong background magnetic field.  
In order to explore the flow compressibilty effect over the particle dynamics 
we performed different numerical experiments: an incompressible case,
and two weak compressible cases with Mach number $M=0.1$ and $M=0.25$. 
We analyze the behavior of protons 
and electrons in those turbulent fields, which are well known to form 
aligned current sheets in the direction of the guide magnetic field.
What we call protons and electrons are test particles
  with scales comparable to (for protons) and much smaller than (for
electrons) the dissipative scale of MHD turbulence, maintaining
the correct mass ratio $m_e/m_i$. For these test particles
we show that compressibility enhances the efficiency of proton acceleration, and 
that the energization is caused by perpendicular electric fields generated between 
currents sheets. On the other hand, electrons remain magnetized and display an 
almost adiabatic motion, with no effect of compressibility observed.
Another set of numerical experiments takes into account two fluid
modifications, namely electric field due to Hall effect and electron pressure gradient.  We 
show that the electron pressure has an important contribution to electron acceleration 
allowing highly parallel energization.  In contrast, no significant
effect of these additional terms is observed for the protons.
\end{abstract}

\maketitle

%

\section{\label{sec:level1}INTRODUCTION:}
Turbulence is an ubiquitous phenomenon in many astrophysical 
environments, in which a wide variety of temporal and spatial scales
are involved. This is the case of the solar wind or the intellestar
medium where the energy is transferred from large to small kinetic
scales where the energy is dissipated. In the macroscopic
  description of a plasma, magnetohydrodynamics (MHD) turbulence
is the result of the  nonlinear interaction of fluctuations of the
velocity and magnetic fields, leading to a spatial intermittency that
is associated with coherent structures, and where the dissipation is
concentrated in strong gradient regions that impact on the heating,
transport and particle acceleration in plasmas \cite{M1}.

The efficiency of MHD turbulence to accelerate charged particles 
and its importance in space 
physics has been reported by many different authors\cite{F1,L1,M2}, 
but the great variety of 
scales involved in turbulence and the particle dynamics
makes this a challenging problem.
On long timescales (large eddy turnover times) dynamics is
governed by stochastic acceleration, and momentum diffusion is
the main acceleration 
mechanism which has been mainly applied for cosmic-ray energization 
studies and frequently
addressed by quasi-linear theory (QLT)\cite{S1,CH1,Lange1}. 
In diffusion studies 
MHD turbulence is commonly represented as a random 
collection of waves, and that 
representation lacks coherent structures that have an important role 
at particle scales\cite{Vlahos}

Dmitruk et al 2004\cite{PD1}, using test particle simulations in static electromagnetic fields obtained 
from direct numerical simulation (DNS) of the MHD equations, showed that particle 
energization at dissipation scales is due to current sheets, and the acceleration
mechanism depends on the particle gyroradii. By static electromagnetic fields, here we mean
that the fields are dynamically computed in a turbulent and self-consistent MHD simulation,
and then a snapshot is extracted and the fields are frozen to compute particle trajectories
and acceleration.

Using a more sophisticated model, but still using static turbulent 
electromagnetic fields, Dalena et al 2012\cite{Dalena2012} showed essentially the same 
results. Electrons initially moving with Alfv\'en velocity experience parallel 
(to the guide magnetic field) acceleration by parallel electric fields inside current 
sheet chanels. On the other hand, protons are accelerated in a two stage process: 
Initially they are parallelly accelerated and gain substantial energy in a short time. Then,
when the proton gyroradius becomes comparable to the current sheet
thickness, protons are accelerated perpendicular to the guide field.  
\newpage
Effects of compressible MHD on particle energization has been reported 
in diffusion studies\cite{Chandran2003,CHO1}, where supersonic
turbulence was considered. There are also reports of test particle pitch angle scattering in 
compressible MHD turbulence\cite{Lynn2013} considering 
second order Fermi acceleration by weak compressible MHD running simultaneously
the test particles and MHD fields, and imposing a scattering rate.
It was found that compressibility is important to produce non-thermal particles. 
Additionally, there are other
studies where test particles and fields are simultaneously advanced in time.
Weidl et al 2015\cite{Weidl2015} and Teaca et al. 2014\cite{Bogdan2014} 
used an incompressible MHD model, analyzing the effect of the 
correlation between magnetic and velocity fields 
on pitch-angle scattering and particle 
acceleration. They found that imbalanced turbulence (nonzero cross-helicity in 
the system) reduces the particle acceleration and also 
the pitch angle scattering.

In the present work we are interested in the compressibility effect 
on particle acceleration by coherent structures in static
electromagnetic fields stemming from a direct numerical simulation of
the MHD equations, and in the identification of the fields which
accelerate the particles. We analyze the particle behavior for three 
different situations: an incompressible case, and two weakly compressible cases with 
differing values of the sonic Mach number. 
We also consider the effect of the Hall current and of
  electron pressure in the acceleration. The organization of this
  paper is as follows:  In section 2 we describe the model employed in
  our investigation, the equations and properties of turbulent MHD
  fields, and the test particle model including the parameters that
  correlate particles and fields. In sections 3 and 4 we show the
  properties of proton and electron dynamics. Finally, in section 5 we
  discuss our findings and present our conclusions.

\section{\label{sec:level2}MODELS:}
The macroscopic description of a plasma adopted here
is the system of the three-dimensional compressible MHD equations: the continuity
(density) equation, the equation of motion, the magnetic field induction equation, and
the equation of state. These are Eqs. (1-4) respectively, which involve fluctuations  of the
velocity field $\textbf{u}$, magnetic field $\textbf{b}$, and density $\rho$. We assume
a large-scale background magnetic field $B_0$ in the $z$-direction, so that the total magnetic
field is $\mathbf{B = B_0 + b}$

\begin{equation}
 \frac{\partial \rho}{\partial t} + \nabla \cdot (\textbf{u}\rho) = 0,
\end{equation}

\begin{equation}
 \frac{\partial \textbf{u}}{\partial t} + \textbf{u} \cdot \nabla \textbf{u} = - \frac{\nabla p}{\rho} + \frac{\textbf{J} \times \textbf{B}}{4\pi\rho} 
 + \nu \left( \nabla^2 \textbf{u} +    \frac{\nabla \nabla \cdot \textbf{u} }{3} \right),
\end{equation}

\begin{equation}
\frac{\partial \textbf{B}}{\partial t} = \nabla \times (\textbf{u} \times \textbf{B}) + \eta \nabla^2 \textbf{B},
\end{equation}

\begin{equation}
 \frac{p}{\rho^{\gamma}} = {\rm constant}.
\end{equation}

\noindent Here $p$ is the pressure, $\nu$ the viscosity, $\eta$ the magnetic 
diffusivity, and $\textbf{J}=\nabla \times \textbf{B} $ is the current density. 
We assume a polytropic 
equation of state $p/p_0= (\rho/\rho_0)^{\gamma}$, with $\gamma=5/3$, where
$p_0$ and $\rho_0$ are respectively the equilibrium (reference) pressure and density.  
We consider two weak compressible cases with Mach number
($M= \sqrt{\gamma p_0/\rho_0}$) equal to $M=0.1$ and $M=0,25$. Additionally, 
in order to have a reference to measure the effect of compressibility on particle 
acceleration, we consider an
incompressible case (with $\nabla \cdot \textbf{u} = 0$ and $\rho=$
a uniform constant).

The magnetic and velocity fields are here expressed in Alfv\'en speed units; 
a characteristic plasma velocity is given by the parallel Alfv\'en wave velocity
along the mean magnetic field $v_A = B_0/\sqrt{4\pi\rho_0}$. An Alfven speed based on field 
fluctuations can also be defined as $v_0=\sqrt{\left<b^2\right>/4\pi\rho_0}$. 
The ratio of the fluctuating to the mean magnetic field is
  $\left< b \right> /B_0\approx0.1$.
The ratio of fluid equilibrium pressure $p_0$ to 
magnetic pressure $B_0^2$,
the so-called $\beta$ of the plasma, is $\beta = p_0/B_0^2=1/(M^2 B_0^2)
= 0.25$.
We take $v_0$ as a unit for velocity and magnetic field
  fluctuations. We use the isotropic MHD turbulence correlation length
  $L$ as a characteristic length (also called the energy containing
  scale), defined as $L = 2\pi \int (E(k)/k) dk / \int E(k) dk$ where
  $E(k)$ is the energy at wavenumber $k$. The value of this scale for
  our simulations is $L=1.3$, as compared to the box size $L_{box}=2\pi$.
  The unit timescale $t_0$, also called eddy turnover time, is derived
  from the unit length and the fluctuation Alfven speed $t_0=L/v_0$. 
  We note that our simulations do not have exact equipartition between 
  magnetic and kinetic energy, this ratio being $E_m/E_k \approx 0.8$.
  The initial magnetic and velocity field fluctuations populate an
  annulus in Fourier k-space defined by a range of wavenumbers with 
  $3\leq k \leq4$, with constant amplitudes and random phases.

\begin{figure*}[<t>]
\begin{center}
{\includegraphics[width = 0.42\textwidth]{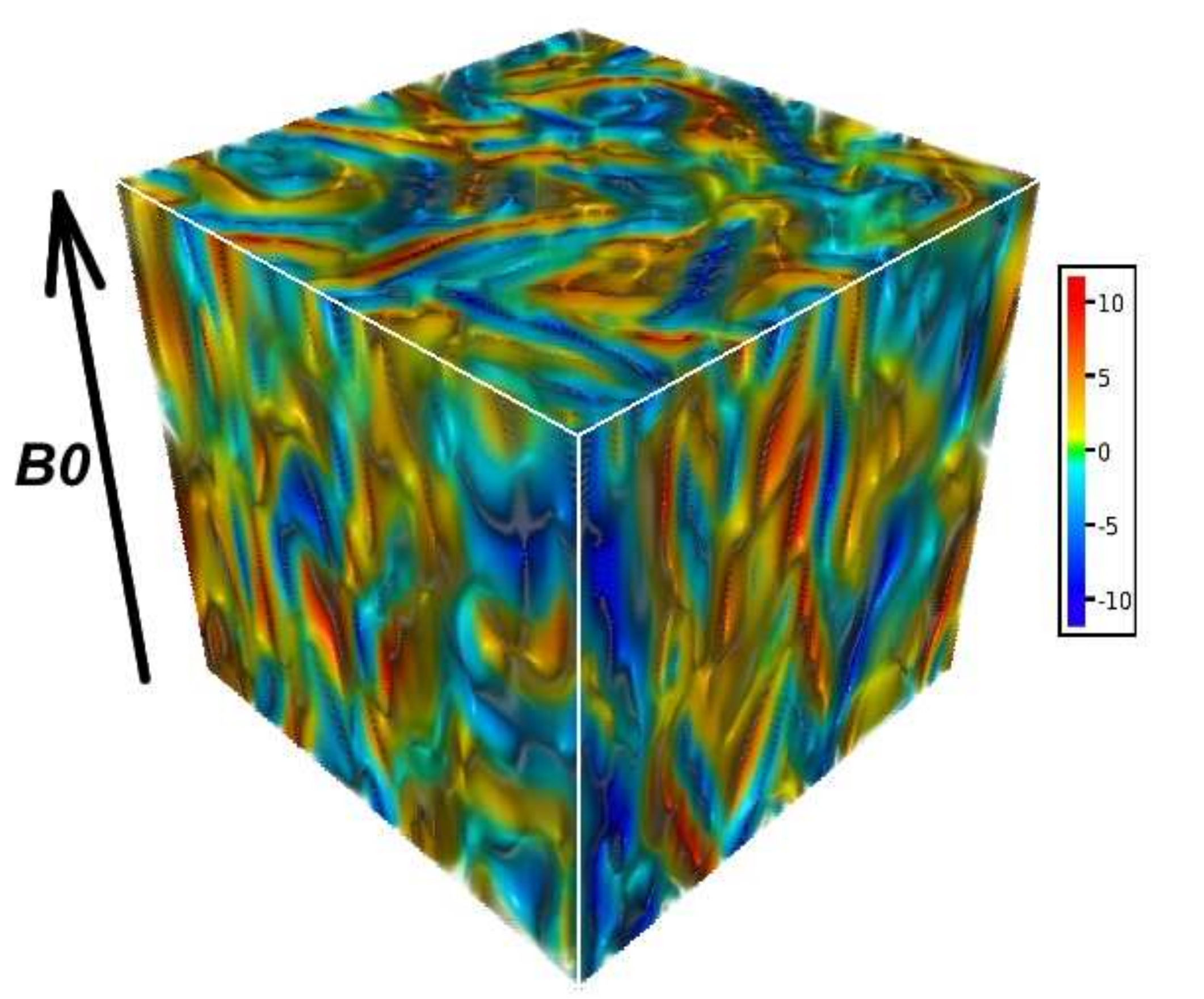}}
{\includegraphics[width = 0.42\textwidth]{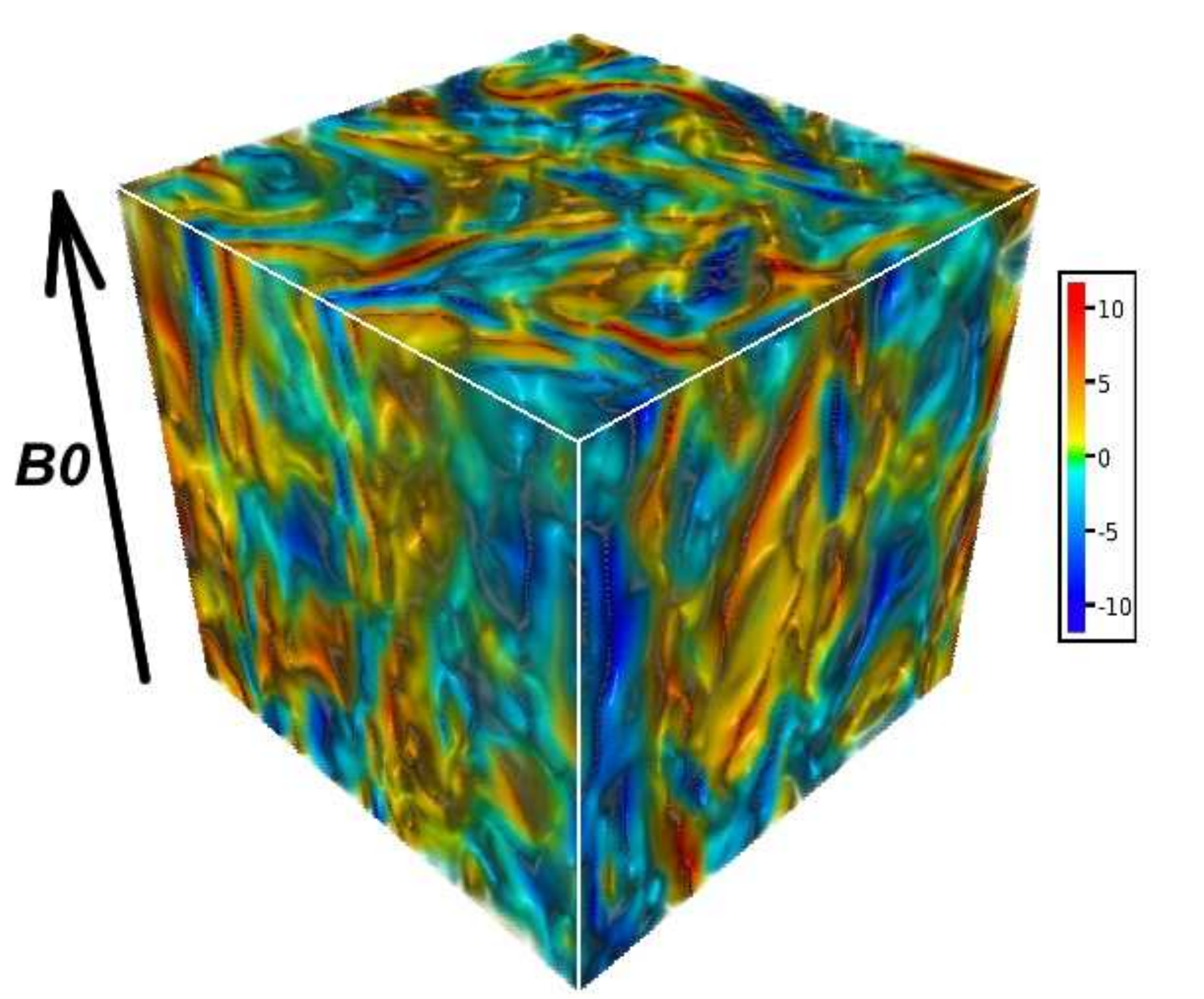}}
\caption{Three-dimensional view of the parallel current density $J_z(x,y,z)$. (Left) 
Incompressible and (Right) Compressible case with Mach number $M=0.25$ at $t/t_0 =2$.}
\end{center}
\end{figure*}
The MHD equations are solved numerically using 
a Fourier pseudospectral method with periodic 
boundary conditions in a cube of size  $L_{box}$; 
this scheme ensures exact energy conservation for the continuous time spatially discrete
equations\cite{ghost}. The discrete time integration is done with a high-order Runge-Kutta
method, and a resolution of ($256^3$) Fourier modes is used. For the kinematic Reynolds
number $R=v_0L/\nu$ and the magnetic Reynolds number $R_m=v_0L/\eta$, 
we take $R=R_m= 1000$, which are limited here by the available
spatial resolution.

When the turbulence is fully-developed a broad range of 
scales develops, from the outer scale $L$ to 
the Kolmogorov dissipation scale $l_d=(\nu^3/\epsilon_d)^{1/4}$, 
with $\epsilon_d$ the average rate of 
energy dissipation. For the simulations it is $\l_d\approx 1/32$. 
We then employ a snapshot of this turbulent MHD state in which to
evolve the test particles. The behavior of a test particle in an
electromagnetic field is described by the nonrelativistic particle
equation of motion:

\begin{equation}
  \frac{d\textbf{v}}{dt} = \alpha(\textbf{E} + \textbf{v} \times \textbf{B}), \ \ \ \  \frac{d\textbf{r}}{dt} = \textbf{v}.
\end{equation}

The nondimensional electric field \textbf{E} is obtained from Ohm's law 
normalized with $E_0= v_0 B_0/c$ as follows:

\begin{equation} 
 \textbf{E} =  -\textbf{u}  \times \textbf{B} + \frac{\textbf{J}}{R_m}. 
\end{equation}

Finally the adimensional parameter $\alpha$ relates particles and MHD field parameters:
\begin{equation}
\alpha=Z\frac{m_p}{m}\frac{L}{\rho_{ii}},
\end{equation}
where $\rho_{ii}$ is the proton inertial length given by
$\rho_{ii}=m_pc/(e\sqrt{4\pi\rho_0})$, $m$ is the mass of the
particle, $m_p$ is the mass of the proton, and $Z$ is the atomic
number ($Z=1$ for protons and electrons). The inverse $1/\alpha$
represents the nominal gyroradius, in units of $L$ and with
velocity $v_0$ and measures the range of scales involved in the system
(from the outer scale of turbulence to the particle gyroradius). One
could expect a value $\alpha \gg 1$ specially for space physics and
astrophysical plasmas. This represent a huge computational challenge
due to  numerical limitations. As stated above, we consider here a
dissipation length scale $l_d\approx 1/32$, which is also of the order
of the current sheet thickness.

In the fixed MHD turbulence state, $10000$ test particles are 
randomly distributed 
in the computational box and the equation of motion of particles 
subject to the MHD
electromagnetic field are solved using a second-order Runge-Kutta method. 
Furthermore, we use high order spline interpolation to compute the field values on each
particle position.

Particles are initialized with a Gaussian velocity distribution function with a 
root mean square (rms) value of the order of the Alfven velocity. It is well known that the 
particle gyroradius has a significant influence 
on acceleration, and our aim in this paper is to explore the compressibility effect on 
acceleration of large gyroradius 
and small gyroradius particles. 
In the next section we show two different compressible cases 
with Mach number $M=0.25$ and $M=0.1$, as well as an incompressible
case. In all cases the mean magnetic field is set to $B_0=10$. We
present the behavior of protons with a nominal (speed $v_0$)
gyroradius $1/32$, and electrons  ($m_e=m_p/1836$) with nominal
gyroradius $1/58752$.

\section{\label{sec:level3}FLOW COMPRESSIBILITY EFFECTS (FCE):}
In Figure 1 a three-dimensional view of the z-component of the current
density $J_z(x,y,z)$ is shown at $t=2.5t_0$ for the incompressible case
and a compressible case with $M=0.25$. It is observed that current
sheets are aligned in the direction of the guide magnetic field. 
It can also be seen that in both cases the structures are similar,
but more corrugated in the compressible case and smoother in the
incompressible one. It is worth mentioning that we used the same 
initial conditions for all the simulations. Coherent structures like
these show the natural tendency of the MHD equations to develop strong
gradients leading to many reconection zones, which is well known to be
one of the mechanisms behind charged particle acceleration.
Figure 2 shows the spectrum of kinetic (top) and magnetic energy
(bottom) for the  compressible cases with $M=0.25$ and $M=0.1$, and
the incompressible case. In the inertial range there are almost no
differences between the compressible and incompressible energy spectra
for both magnetic and velocity fields, although slightly more energy at 
large scales is observed in the incompressible case. On the other hand, at wavenumbers beyond
the dissipation scale (that is, for $k\geq32$),  an excess of energy is observed as the Mach
number is increased. This feature is more evident for the kinetic energy spectrum
than for the magnetic energy spectrum. Since protons mostly interact with structures
of that size, this can be an important effect on proton acceleration.
\begin{figure}[<h!>]
\begin{center}
\hspace*{0.4cm}{\includegraphics[width = 3.31in]{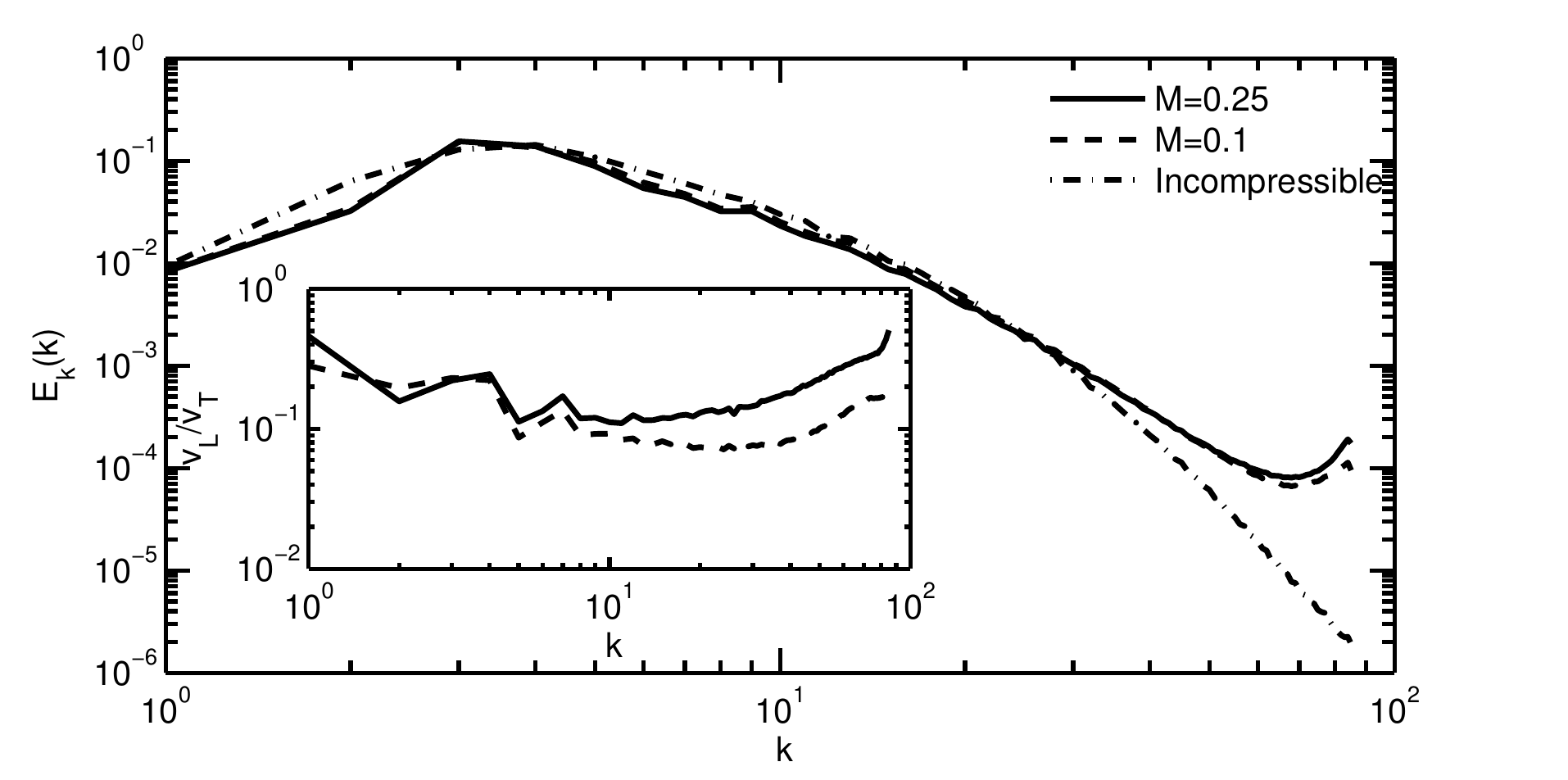}}\\
{\includegraphics[width = 3.05in]{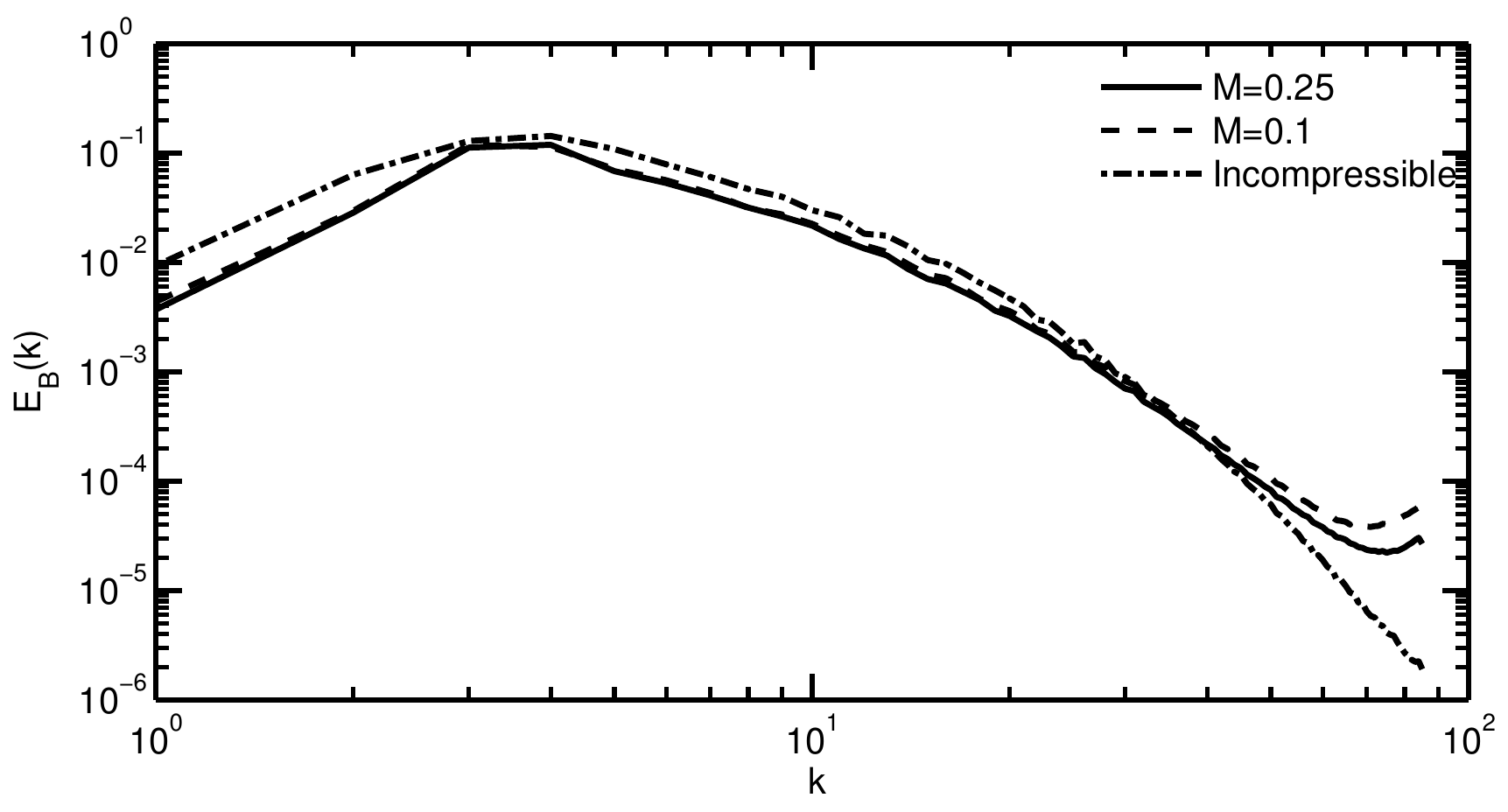}}\\
\caption{(Top) Kinetic energy spectrum for compressible cases with Mach 
numbers $M=0.25$ (solid line), $M=0.1$ (dashed line), and incompressible case 
(dashed-dot line); the inset shows the ratio between solenoidal
and irrotational (compressive) 
components of the velocity field for compressible runs. (Bottom) Magnetic energy spectrum
for the three cases mentioned before, using the same labels.}
\end{center}
\label{mean square velocity}
\end{figure}
In order to explore the importance of compressible effects on MHD fields,
we make a Helmholtz decomposition of the velocity field, presented
in the inset in Fig 2, where $\bf{\hat v_T}({\bf k}) = (\bf I - \hat k\hat k) u({\bf k})$ 
represents the solenoidal (incompressible) part 
and $\bf{\hat v_L}({\bf k}) = \hat u({\bf{k}}) - \hat v_T({\bf k})$ is
the irrotational (compressive) component. It is observed that at high
$k$ the velocity field spectrum is strongly compressible, and that
compression becomes more prominent at higher turbulent Mach number. 
The large $k$ effects in the compressible kinetic spectrum
  may be attributed to the emergence of shock-like structures that enhance the energy in the
  smallest scales, as compared to the incompressible case.

{\it Protons.}
We remark here that what we call ``protons'' are test particles
with gyroradius comparable to the dissipative scale of the MHD turbulence,
although the MHD approximation is only marginally valid at those small
scales.
Figure 3 shows the time evolution for the mean value of
the perpendicular $v_\perp = \sqrt{v_x^2+v_y^2}$ 
(top) and parallel $v_\parallel=v_z$ (bottom) proton velocity, 
relative to $B_0$, for the compressible ($M=0.25$, $M=0.1$) cases
and the incompressible case. 
The typical acceleration process observed in previous studies is
evident, proton are accelerated perpendicularly with respect to $B_0$, 
while they are less accelerated parallely. 

Moreover, the compressibility effect on particle acceleration is clearly observed.
Protons are highly accelerated as compressibility of the fluid increases, 
for both perpendicular and parallel directions.
Acceleration of protons is also observed in the 
incompressible case (see inset plot) but
the value of the velocity reached at the end of the simulation is much lower than in both
compressible cases, even with relatively small values of the Mach number $M$ as the ones
considered here.
\begin{figure}[h!]
\begin{center}
{\includegraphics[width = 3.05in]{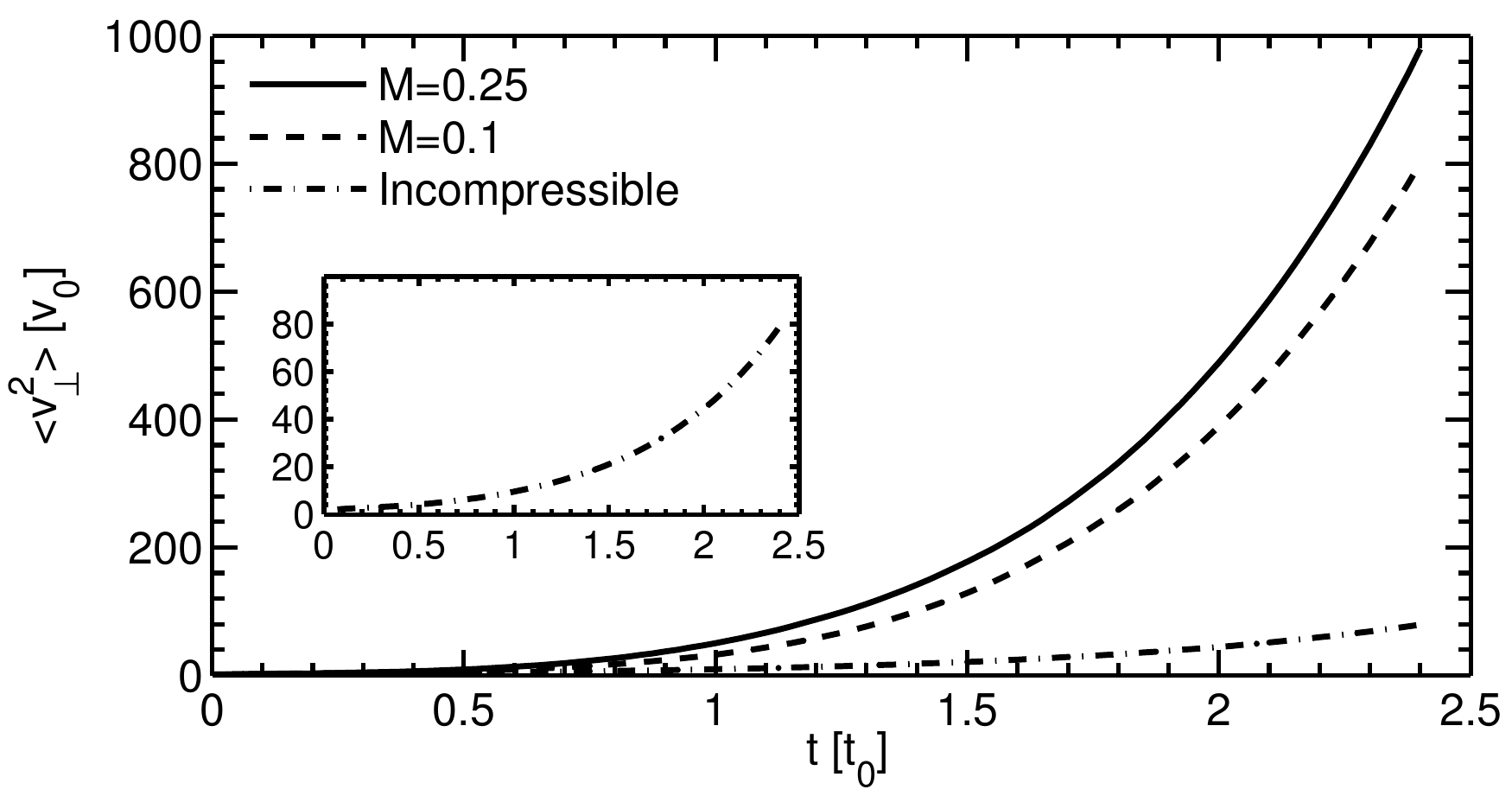}}
{\includegraphics[width = 3.05in]{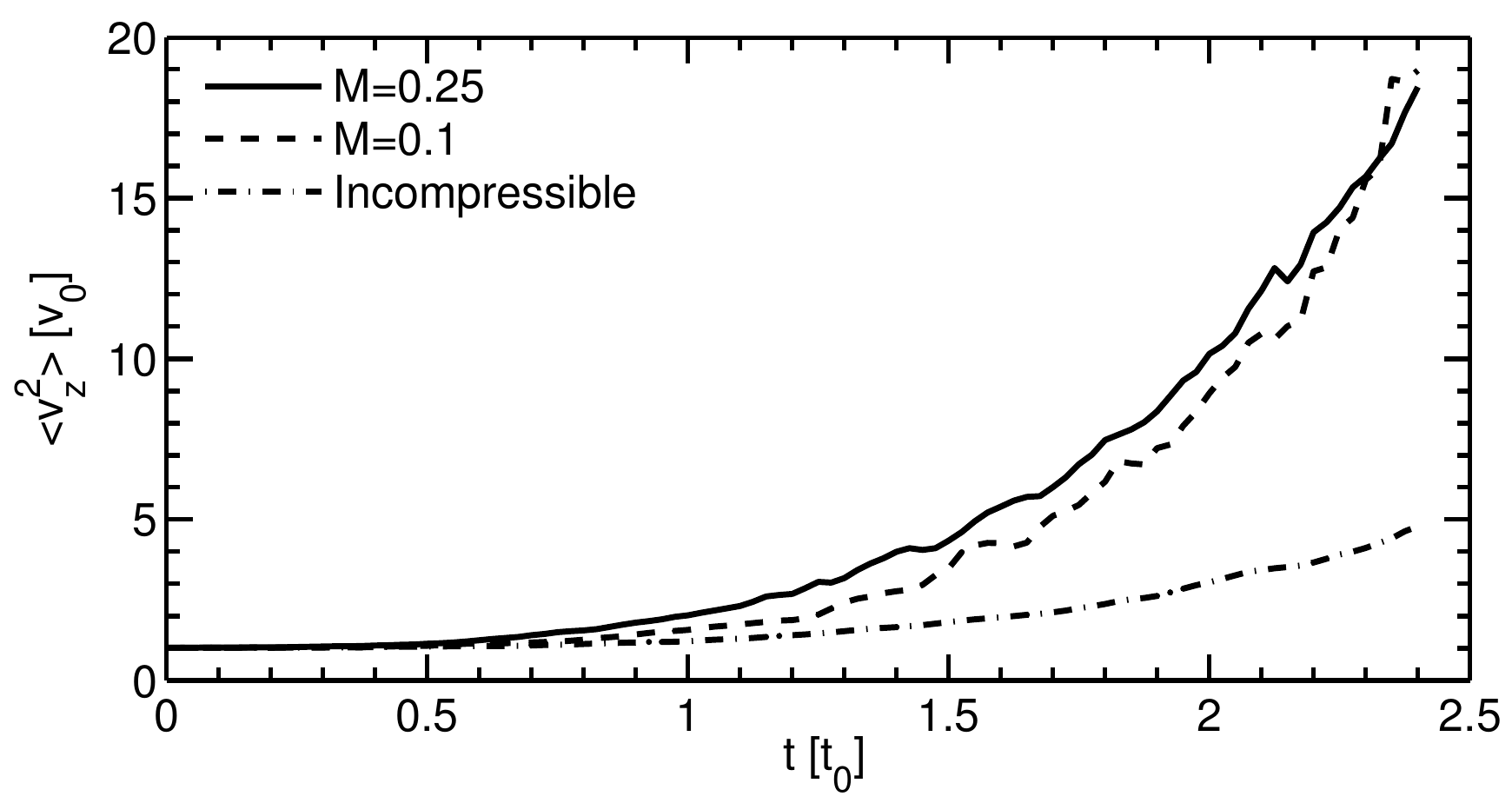}}
\caption{Particle mean square velocity as a function of time: (Top)
Proton perpendicular velocity $v_\perp = \sqrt{v_x^2 + v_y^2}$ 
for two different Mach number
cases, $M=0.25$ (solid line), $M=0.1$ (dashed line), and the
incompressible case (dash-dot line); the inset
shows a detail of the proton perpendicular mean square velocity for the incompressible case.}
(Bottom) Proton parallel velocity $v_\parallel=v_z$ for $M=0.25$, $M=0.1$, 
and incompressible case, with the
same labels for the lines.
\end{center}
\label{mean square velocity}
\end{figure}

\begin{figure}
\begin{center}
{\includegraphics[width = 2.1in]{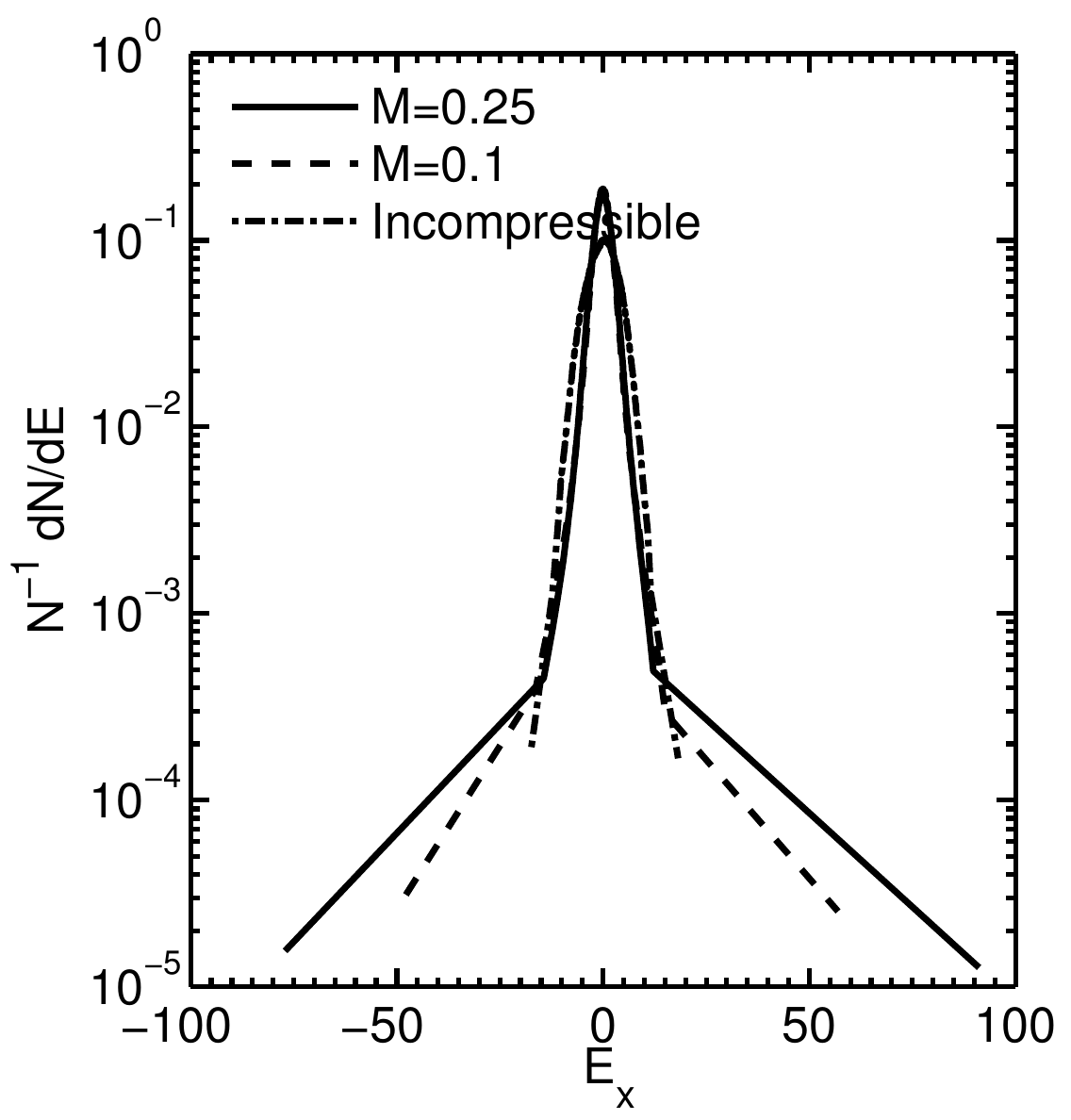}}
{\includegraphics[width = 2.1in]{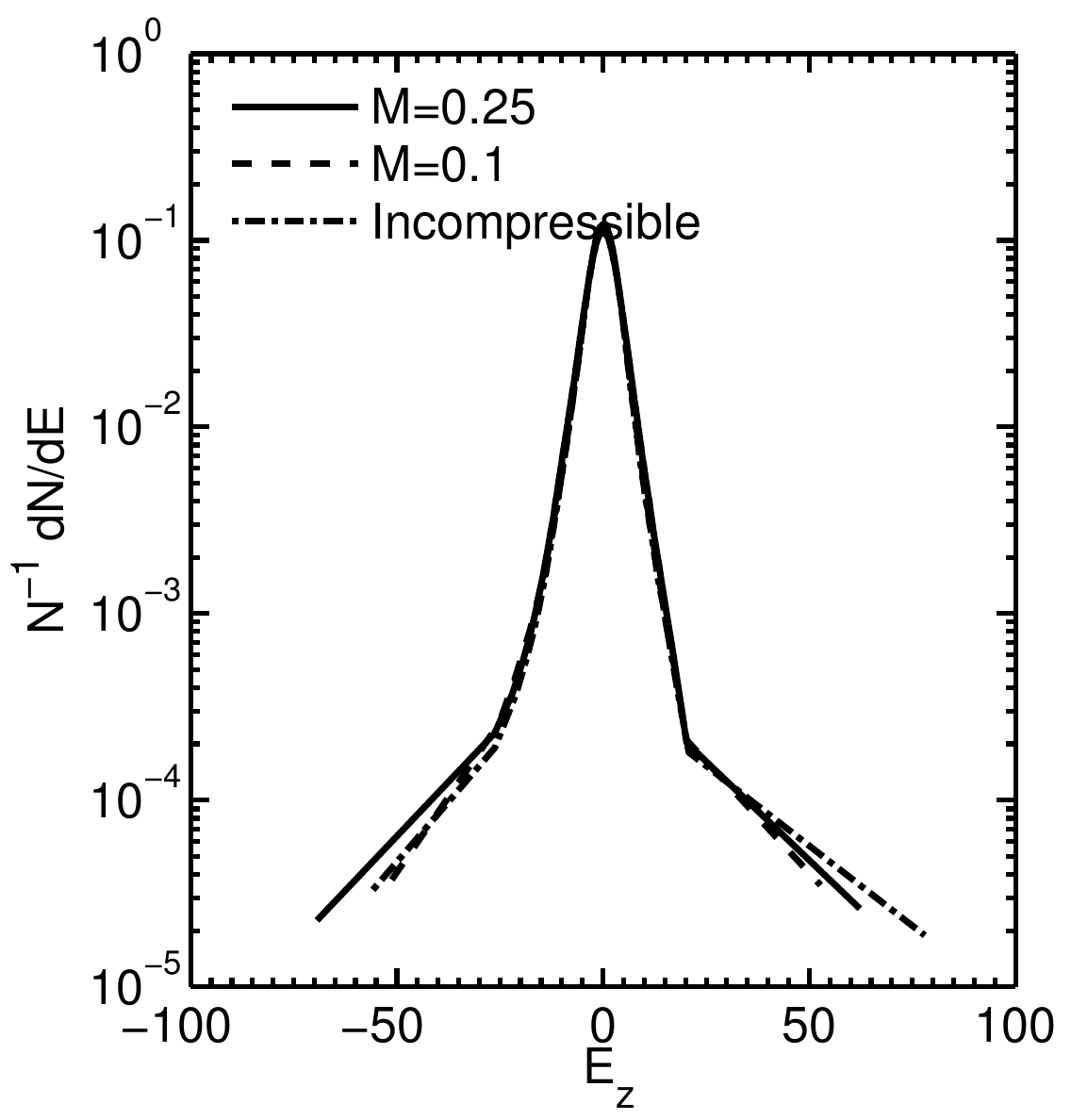}}
\caption{Probability density function of electric field components 
in the simulation. (Top) Perpendicular x-component 
for $M=0.25$ (solid line), $M=0.1$ (dashed line), and incompressible 
(dash-dot line). (Bottom) Parallel z-component of the electric field 
using the same labels.}
\end{center}
\label{mean square velocity}
\end{figure}
Figure 4 shows the probability distribution function (PDF) of the
perpendicular $x$-component (top) and of the parallel $z$-component (bottom) 
of the electric field for the compressible and incompressible cases. The PDF shows that, as 
compression increases, long tails in the distribution arise and higher
values of the perpendicular electric field are achieved. Additionally,
the core part of the distribution function for the incompressible 
case is thicker than for the compressible cases. On the other hand,
the PDF of the parallel electric field shows very little effect of
increasing compressibility. 
In order to better understand the dynamics of protons, in Figure
5 we show the current density $J_z(x,y,z)$ together
with the trajectory of one of the most energetic protons,
for the compressible $M=0.25$ case. The visualization was done using the software
VAPOR\cite{vapor}. It is observed that on the surrounding of the particle trajectory 
there are many current sheets, which contribute to the proton energization.
Figure 6 shows the values of quantities following the trajectory of the
most energetic proton, that is, the most energetic proton is identified
and the values of several quantities along the trajectory of this proton
are obtained: (a) the current density $J_z$, (b) electric field 
components $E_x, E_y, E_z$, (c) proton velocity components $v_x, v_y, v_z$ and (d) root
mean square displacement of the proton. The panels on the left correspond to the
compressible $M=0.25$ case, and the panels on the right correspond to the incompressible case.


\begin{figure}[h!]
\begin{center}
{\includegraphics[width = 2.8in]{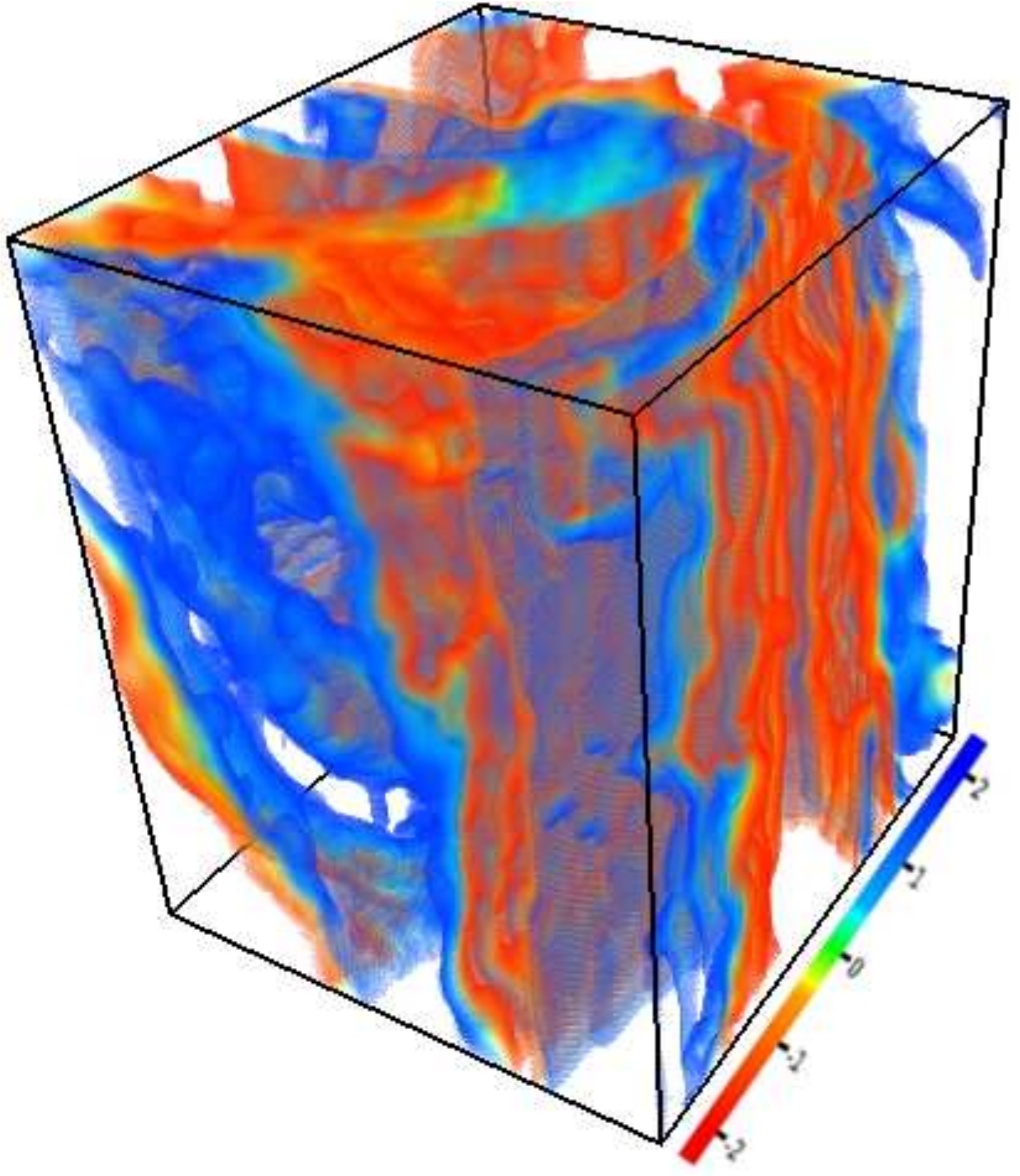}}
{\includegraphics[height=2.5in, width = 3in]{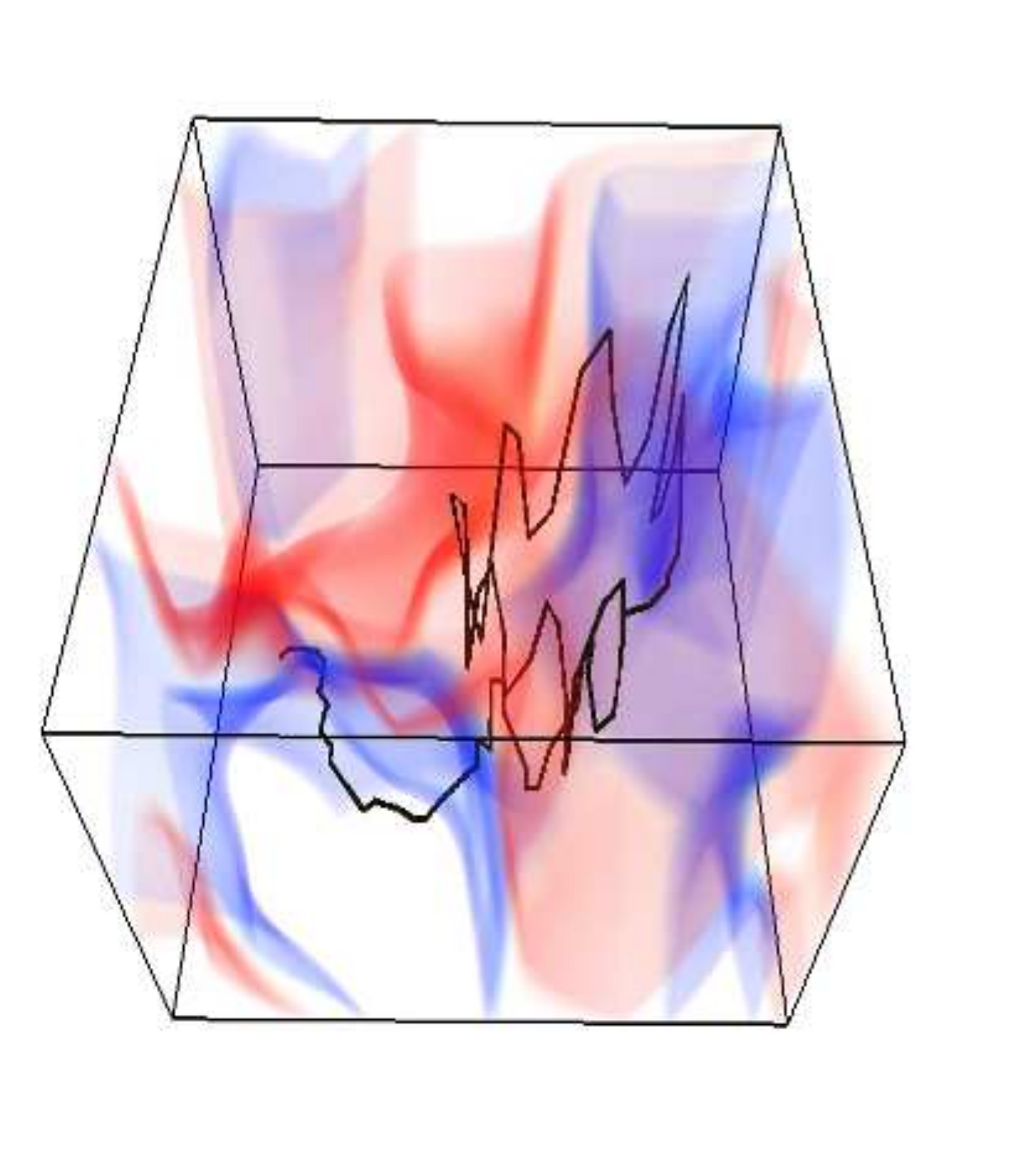}}
\caption{(Top) View of the parallel current density $J_z(x,y,z)$.
(Bottom) Trajectory of one of the most energetic protons; the z-component of the current 
density is shown in the transparent volume rendering.}
\end{center}
\label{mean square velocity}
\end{figure}
It is observed that when there is a change of the sign in the 
current density $J_z$, there is also an increment in the 
perpendicular components of the electric field that 
the particle experiences, and concurrently there is an increment 
of the proton velocity. This situation
is repeatedly observed in time as the energy of the proton increases.

A possible explanation for the change of sign in the current density is 
that the particle is entering and leaving two neighboring 
current sheets with different polarities while experiencing
a strong perpendicular electric field between those current sheets. 
The perpendicular electric field
is stronger as the compression of the fluid increases (this can be noticed
by comparing panels on the left and right of Figure 6).
Consequently, the velocity increment is larger in the compressible case 
than in the incompressible case.
This situation can be 
generalized for many particles in the simulation, 
resulting in the increase of the root mean 
square velocity for the ensemble of particles.

\begin{figure*}[h!]
  \centering
  \begin{tabular}{@{}p{0.45\linewidth}@{\quad}p{0.45\linewidth}@{}}
    \subfigimg[width=\linewidth]{a)}{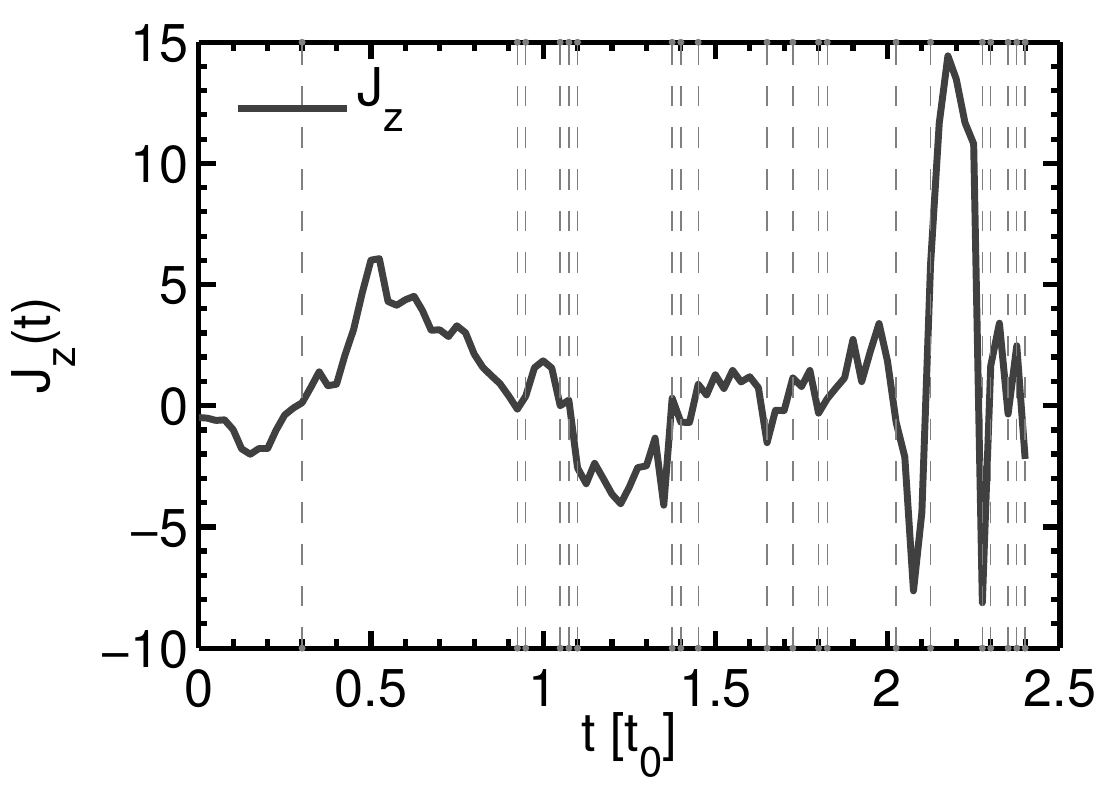} &
    \subfigimg[width=\linewidth]{a)}{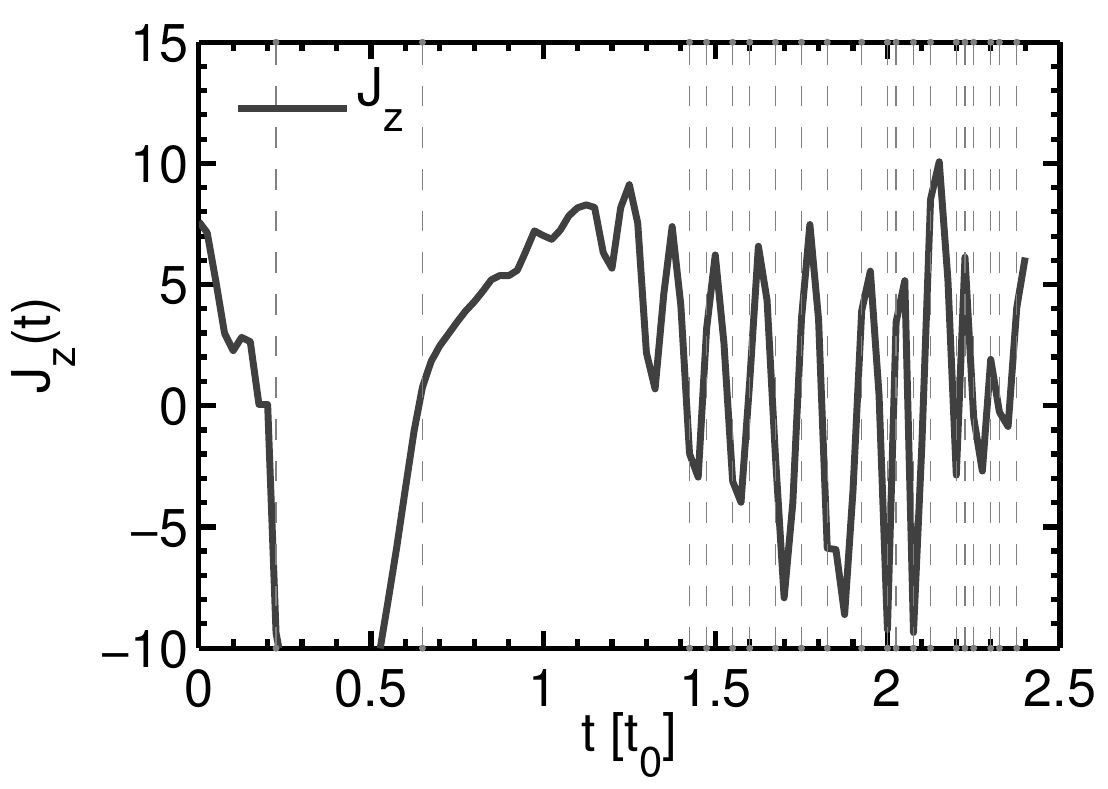} \\
    \subfigimg[width=\linewidth]{b)}{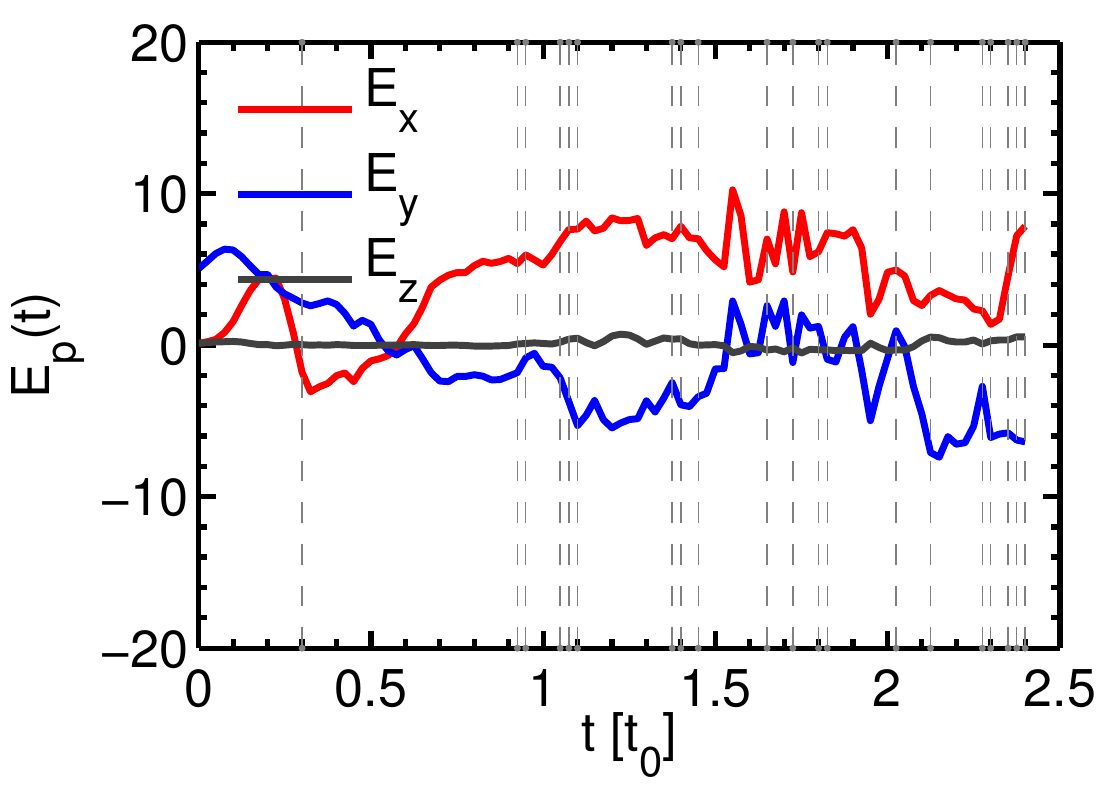} &
    \subfigimg[width=\linewidth]{b)}{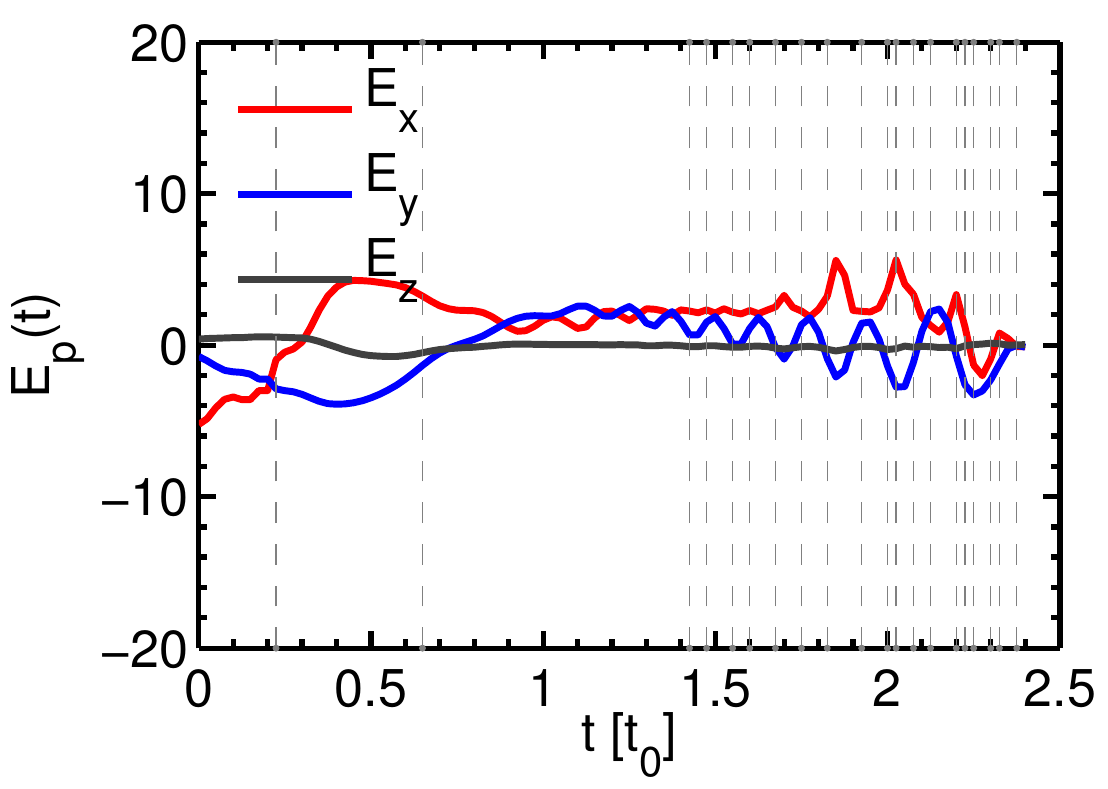} \\
    \subfigimg[width=\linewidth]{c)}{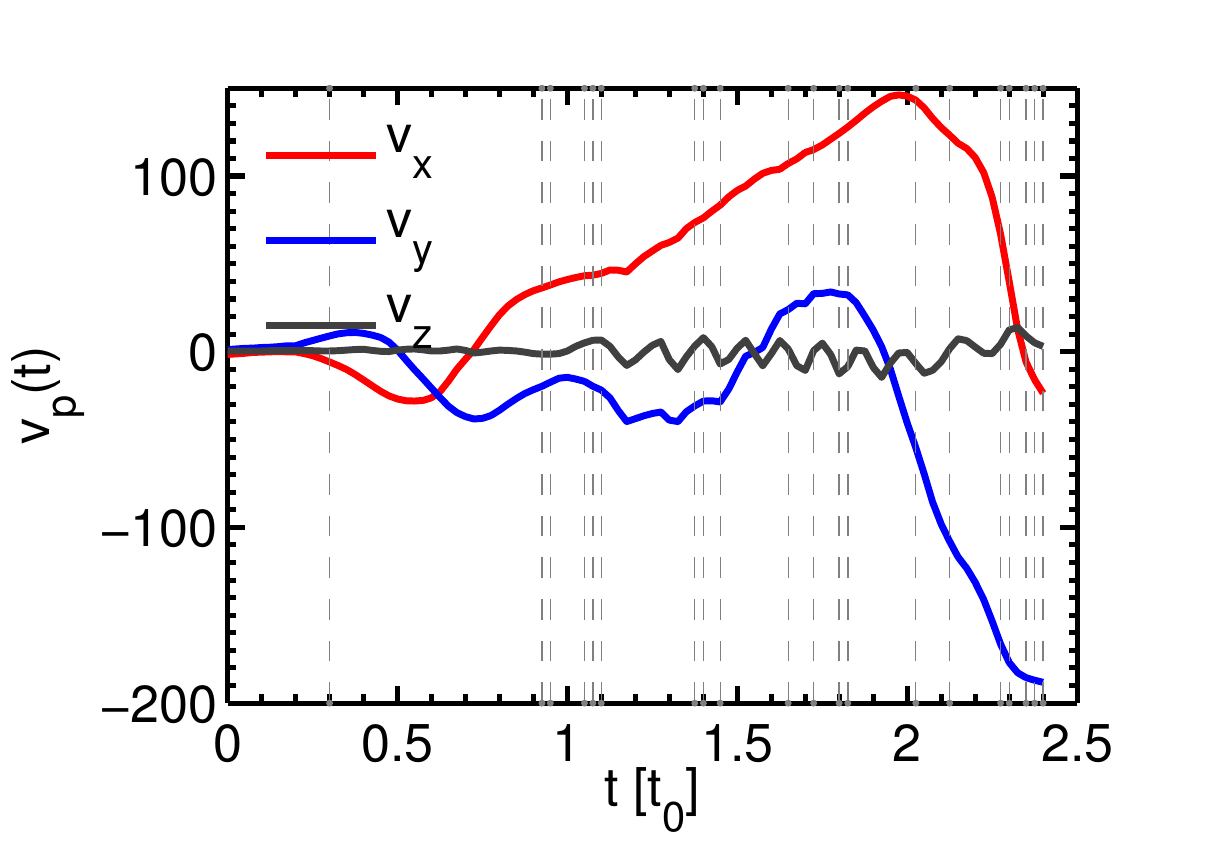} &
    \subfigimg[width=\linewidth]{c)}{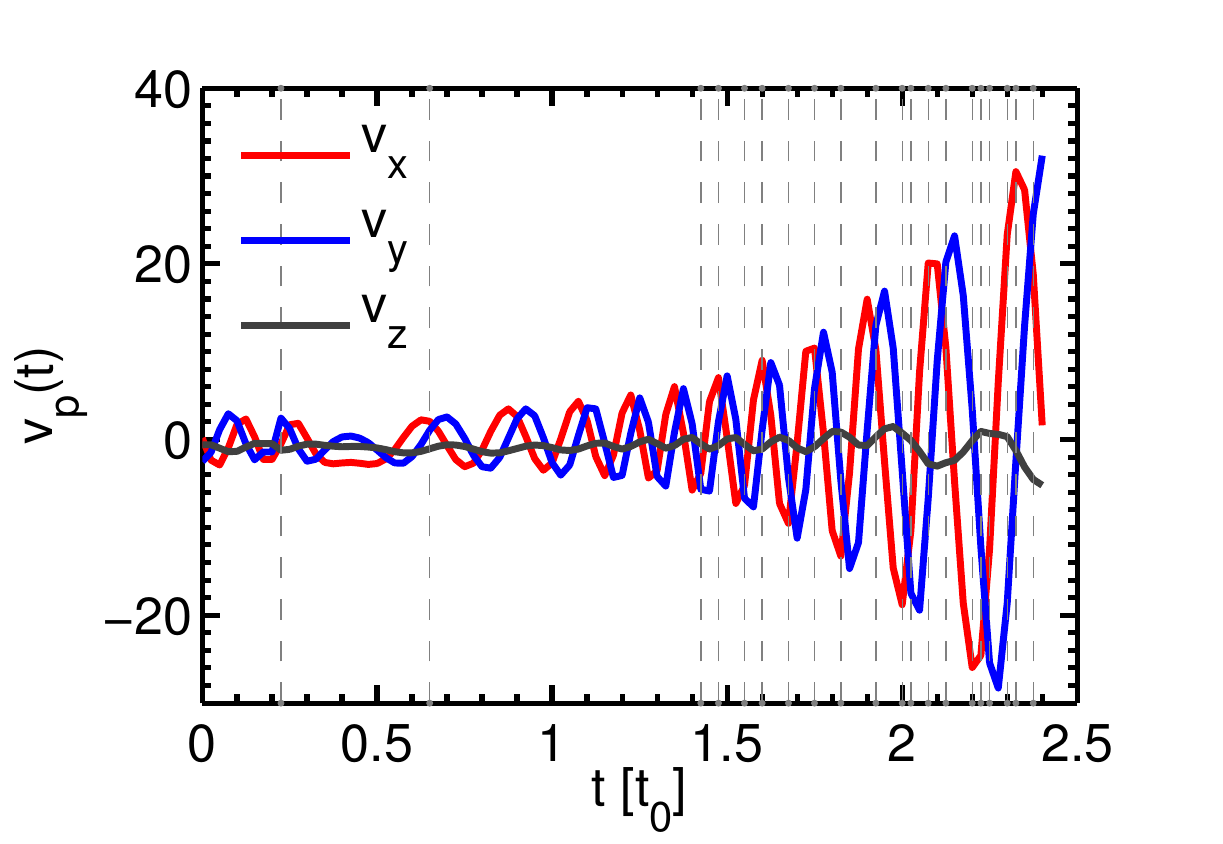} \\
    \subfigimg[width=\linewidth]{d)}{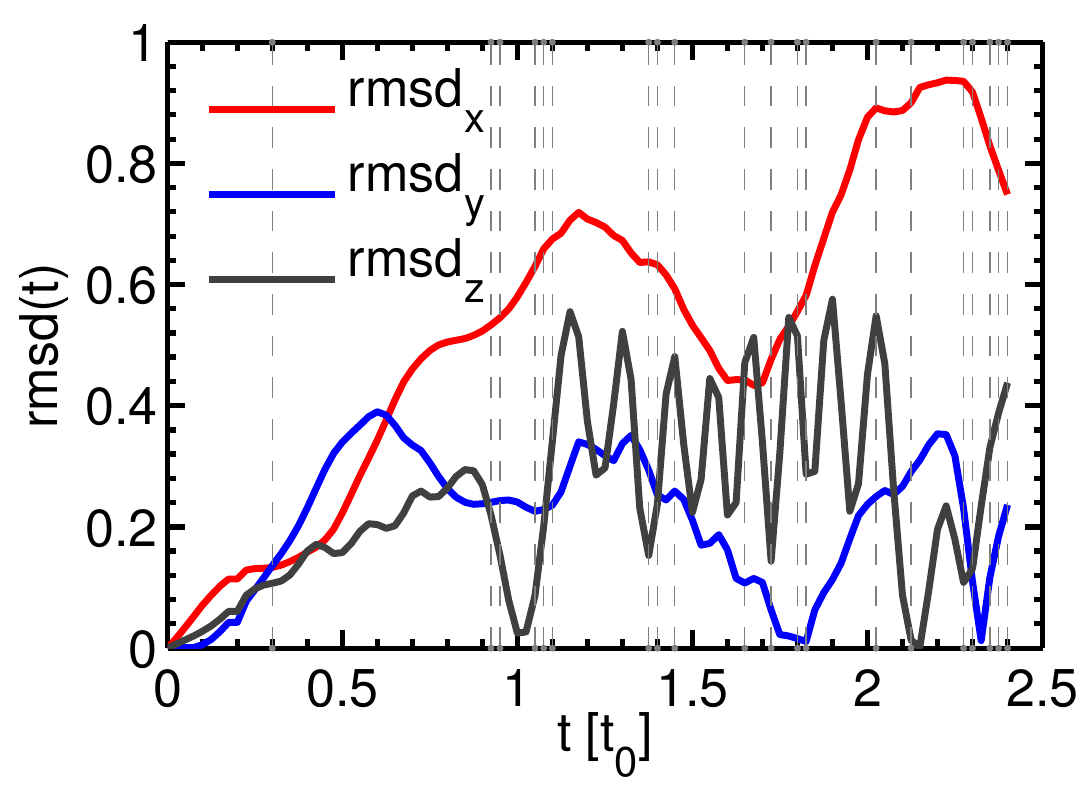} &
    \subfigimg[width=\linewidth]{d)}{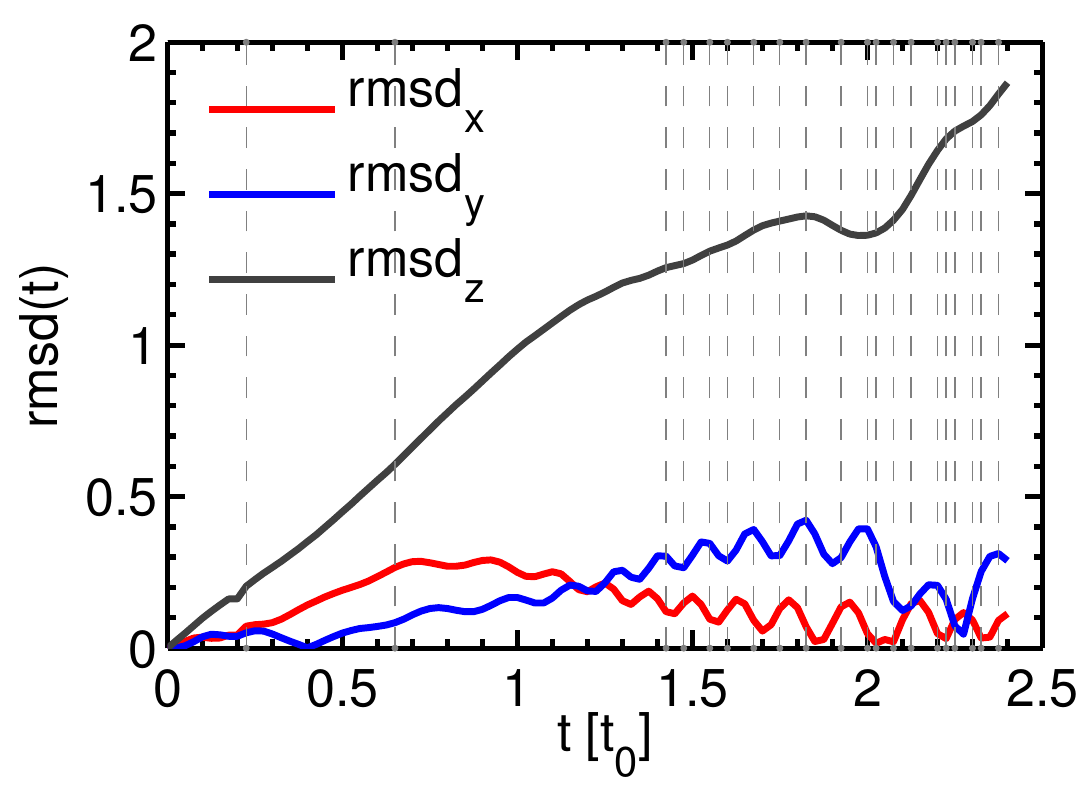}
  \end{tabular}
  \caption{(a) Parallel current density, (b) the three components of the electric field,
  (c) velocity components, and (d) rms displacement as function of time for the most
  energetic particle: (Left) compressible $M=0.25$ case and (Right) incompressible case. The
   gray vertical dashed-lines show the moments when current is reversed.}
\end{figure*}

\clearpage
The reason for greater perpendicular electric field in the compressible cases 
can be understood in terms of the magnetic flux pileup that accompanies the 
interaction of adjacent flux tubes in turbulence\cite{ServidioEA10}. 
While current sheets typically form between interacting flux tubes, 
when the flux tubes are driven together by the turbulent flow, there
is also frequently a magnetic flux pileup near the boundary.
This compression of the magnetic field occurs in the incompressible case as well, 
but clearly can be greater when the material elements themselves are compressible.   
The pileup phenomenon is readily seen to be associated with 
reversal of the electric current density. Furthermore, the parallel magnetic flux increases
due to this compression, requiring a circulation of the perpendicular electric field vector, 
thus setting the scene for betatron acceleration \cite{Dalena2012}.

{\it Electrons.}
What we call ``electrons'' are test particles
with gyroradius much smaller than the dissipative scale of MHD
turbulence. At those scales MHD is not expected to be valid
anymore. However, we maintain the correct ratio of electron to proton
mass, $m_e/m_p=1/1836$. In the next section we discuss other relevant
effects at those scales.

Figure 7 shows the time evolution for the perpendicular (top) 
and $z$-component (bottom) of electron rms velocity for the
compressible ($M=0.25$, $M=0.1$) and incompressible cases. It should
be mentioned that we are showing a short time simulation of electrons
here. This is due to the high computational cost of integrating the
trajectory of electrons in a flow, as electrons require a very small
time step (to represent a physical small gyroradius).  The total time
reached in the electron simulations is of the order of almost 3000 
electron gyroperiods. Electrons present the typical parallel
energization reported in previous works. Besides, there is no evidence
that compression of the MHD fields substantially enhances the electron acceleration,
 as electrons gain almost the same energy regardless the compressible
level of the fluid. Since the gyradius of electrons is smaller than
any of the length scales of structures in the fields, when electrons
find a current sheet they travel along magnetic field lines and there
is not so much difference between compressible and incompressible
cases.

Also, the perpendicular rms velocity shows that electrons are
initially accelerated but quickly exhibit a constant perpendicular
energy. Constant perpendicular energy is consistent with near
conservation of the magnetic moment, which is one of the adiabatic
invariants of charged particle dynamics in a magnetic field. 

It is important to remark that over longer timescales, of the order 
of many turnover times, 
electrons can obtain very high parallel energy, and 
it is likely that the motion will 
no longer be adiabatic.  In that case, 
electrons can reach other regions and interact with 
structures that generate other possible acceleration mechanisms, 
such as those that involve 
pitch angle-scattering, betatron acceleration, etc.

\begin{figure}[<t>]
\begin{center}
{\includegraphics[width = 3.3in]{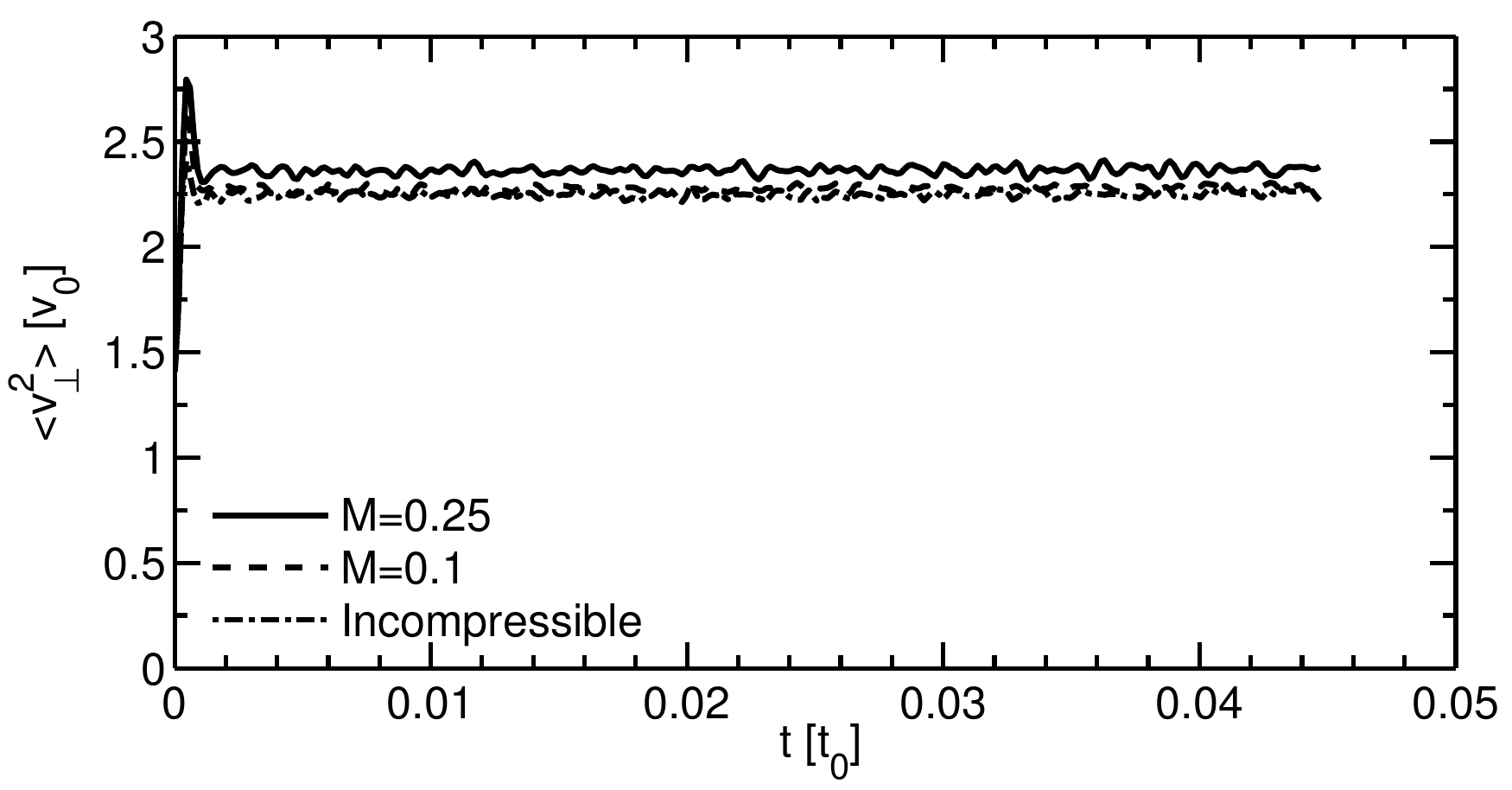}}
{\includegraphics[width = 3.3in]{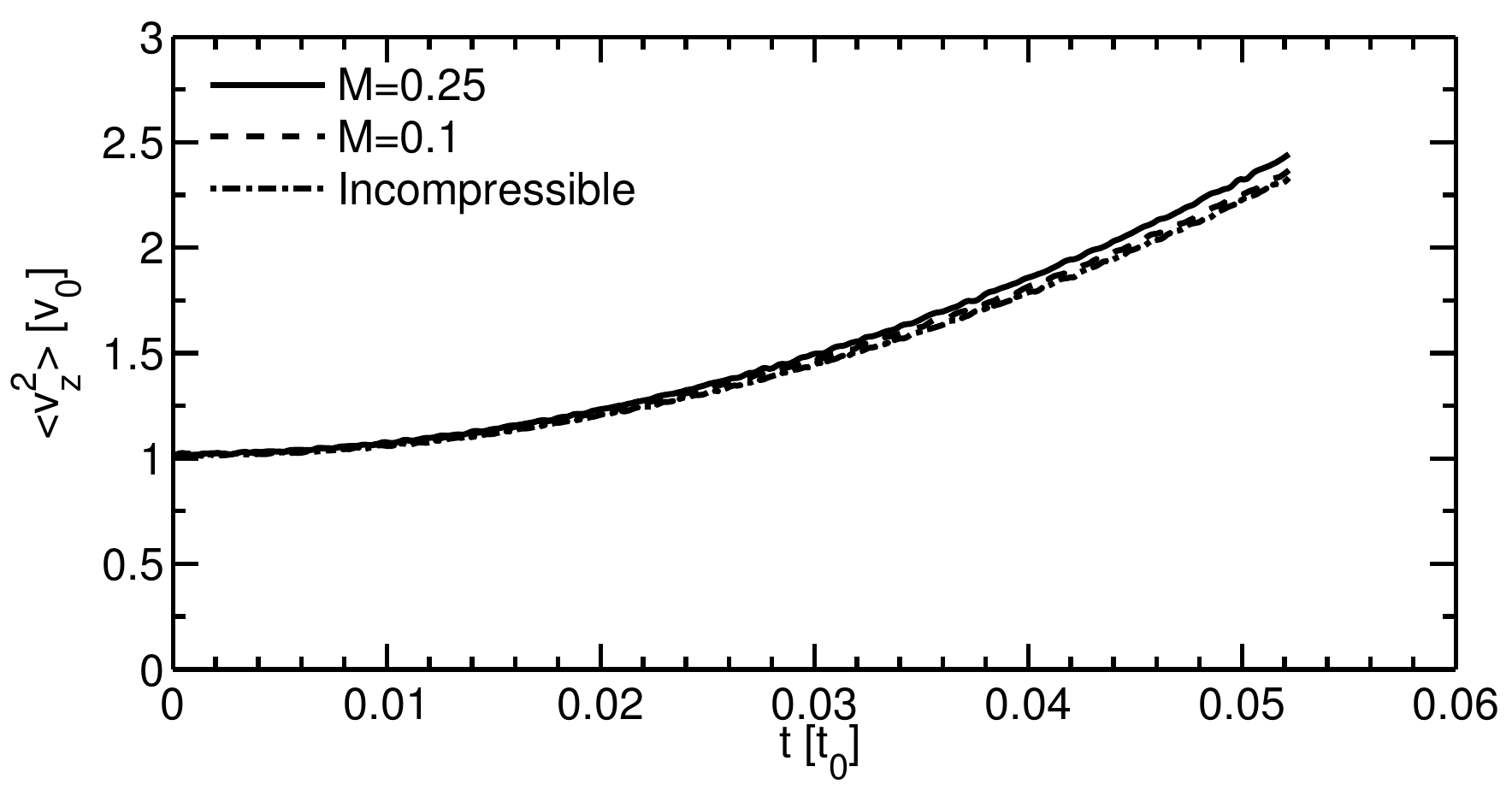}}
\caption{(Top) Time evolution of the perpendicular rms velocity for
electrons, for the compressible cases with $M=0.25$ (solid line), 
$M=0.1$ (dashed line), and the incompressible case (dot-dashed line). (Bottom) Time 
evolution of the parallel rms velocity for electrons, using the same notation.} 
\end{center}
\label{mean square velocity}
\end{figure}

\section{\label{sec:level3}ELECTRON PRESSURE EFFECTS (EPE):}
In this section we consider additional effects in the electric field
which were not taken into account in the previous section. As will be
seen, these effects are important for the electrons but not so for the
protons. Adopting a generalized Ohm's law from a two-fluid plasma description, 
the electric field becomes
\begin{equation} 
 \textbf{E} =  -\textbf{u}  \times \textbf{B} + {\frac{\epsilon}{\rho}}\textbf{J}  \times \textbf{B} - \epsilon \nabla p_e + \frac{\textbf{J}}{R_m}. 
\end{equation}
written in a dimensionless form.

The additional terms as compared to Eq. (6) are the Hall effect
term $\textbf{J} \times \textbf{B}/\rho$ and the electron pressure 
gradient term $\nabla p_e$. The 
dimensionless coefficient $\epsilon$ multiplying terms is 
the Hall parameter:
\begin{equation} 
 \epsilon = \frac{\rho_{ii}}{L}
\end{equation}
which relates the ion inertial length scale with the energy
containing scale. For consistency with the test particles definition
(see Eq. (7)) 
we set the value of the Hall parameter $\epsilon=1/\alpha=1/32$ in
our simulations, where $1/\alpha$ is the nominal gyroradius of the
protons. In the MHD description it is assumed that plasma protons and
  electrons are in thermal equilibrium, i.e, their pressures are
  $p_e=p_i$. Then $p_e = p/2$ with $p=p_e+p_i$ the total pressure.
It is worth mentioning that Dmitruk et al 2006\cite{hall} 
previously analyzed the Hall effect only, not considering electron
pressure effects, and did not see a significant contribution of this
effect in the particles acceleration.

\begin{figure}[ht!]
\begin{center}
{\includegraphics[width = 3.05in]{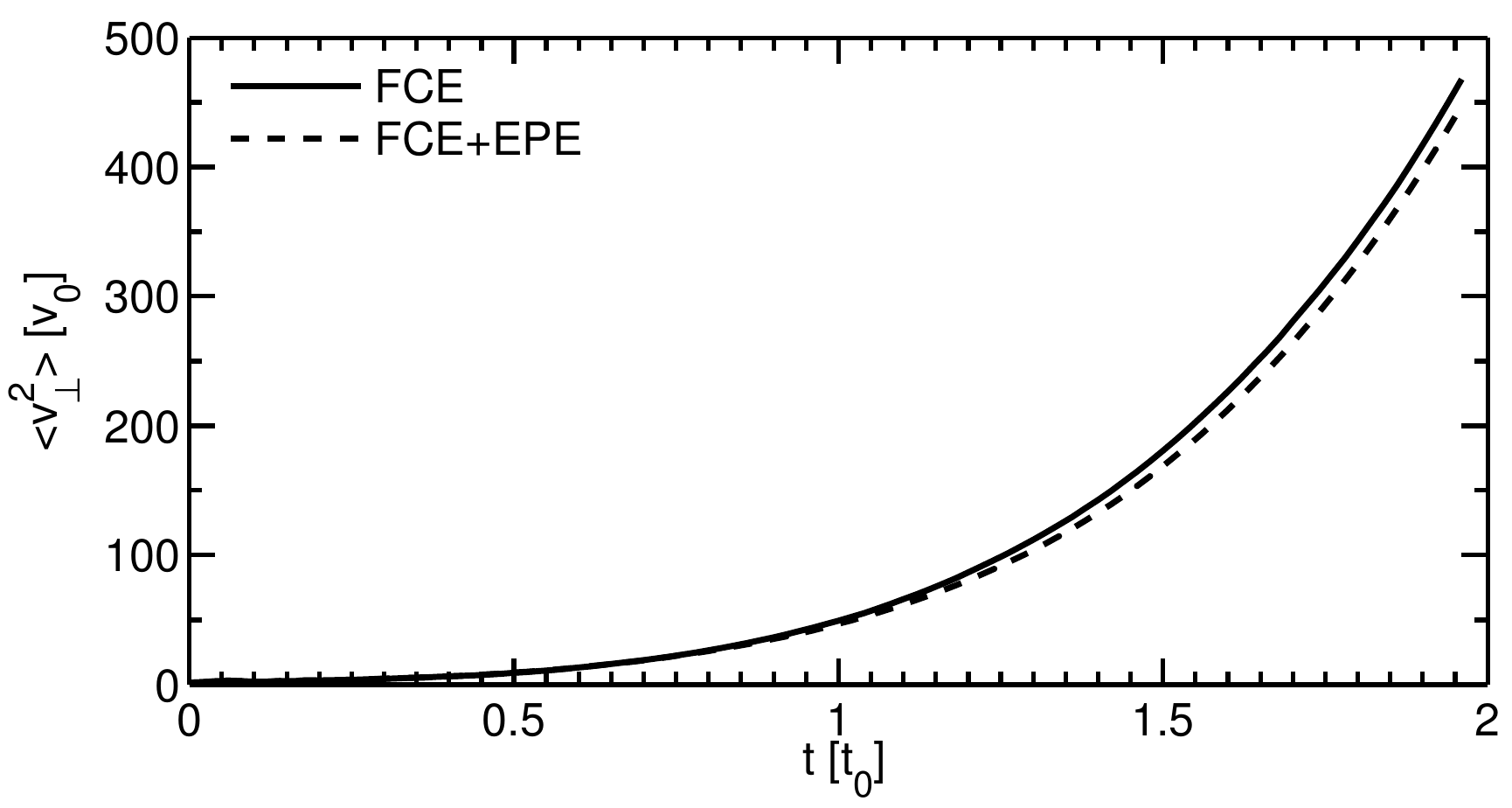}}
{\includegraphics[width = 3.05in]{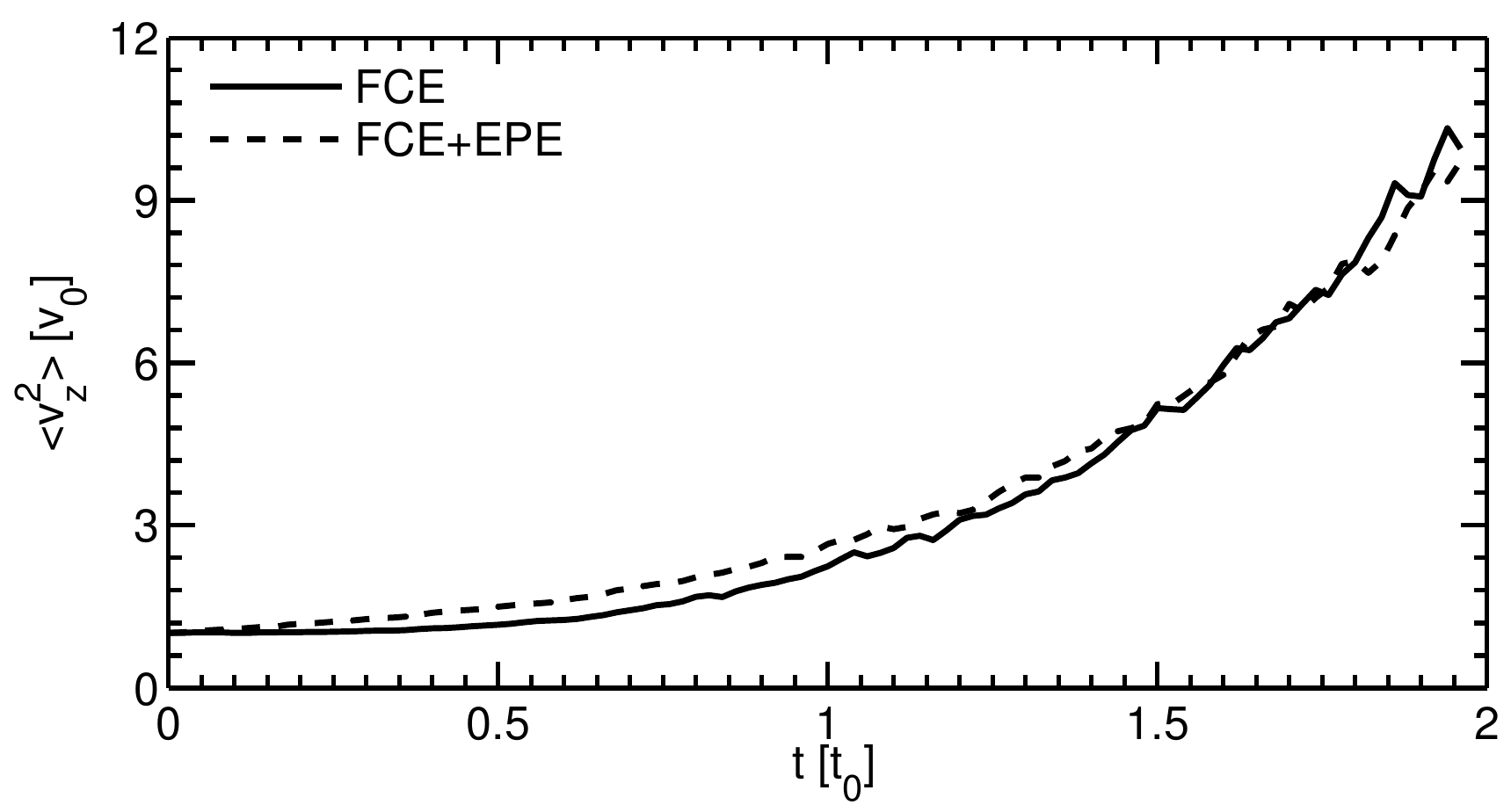}}
\caption{Proton mean square velocity as a function of time considering flow 
compressibility effects (FCE), and considering flow compressibility plus electron 
pressure effects (FCE+EPE): (Top) Proton perpendicular velocity 
$v_\perp = \sqrt{v_x^2 + v_y^2}$ with FCE (solid line) and with
FCE+EPE (dashed line), both for the case with $M=0.25$. 
(Bottom) Proton parallel velocity $v_\parallel=v_z$, with the same
labels for all curves.}
\end{center}
\label{mean square velocity}
\end{figure}

In order to measure the effect of electron pressure on test
  particle energization we compared the proton and electron
  energization for the case with $M=0.25$, taking into account the
  flow compressibility effect (FCE) only, and the flow compressibility
  effect plus the electron pressure effect (FCE+EPE). The results are
  shown in Figures 8 and 9.

Figure 8 shows the perpendicular (top) and parallel rms
  velocity (bottom) for protons. It is observed that no significant
  contribution of EPE occurs for proton energization, and the main
  particle acceleration mechanism remains the interaction with current sheets
  as discussed in the previous section.

\begin{figure}[ht!]
\begin{center}
{\includegraphics[width = 3.05in]{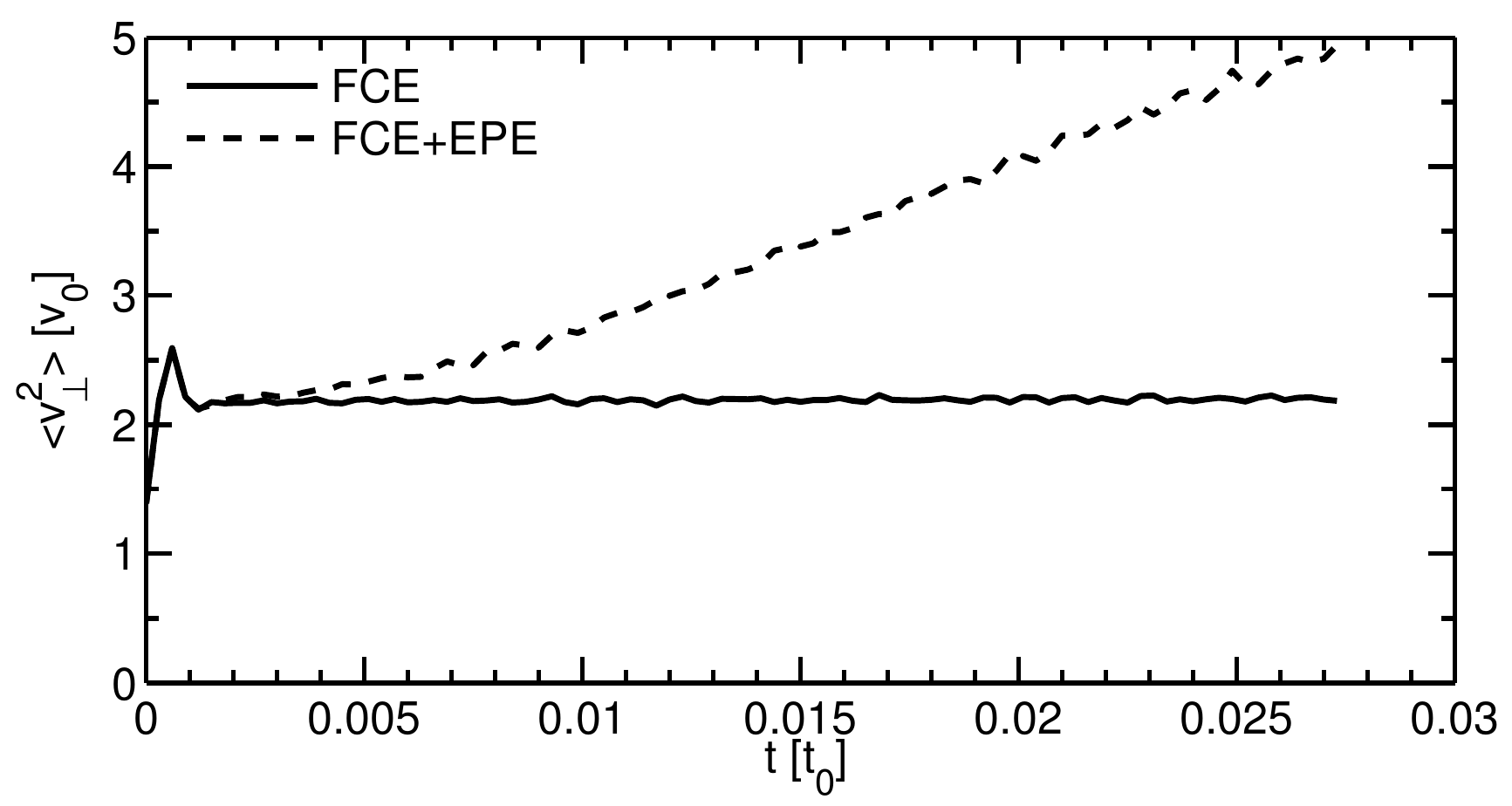}}
{\includegraphics[width = 3.05in]{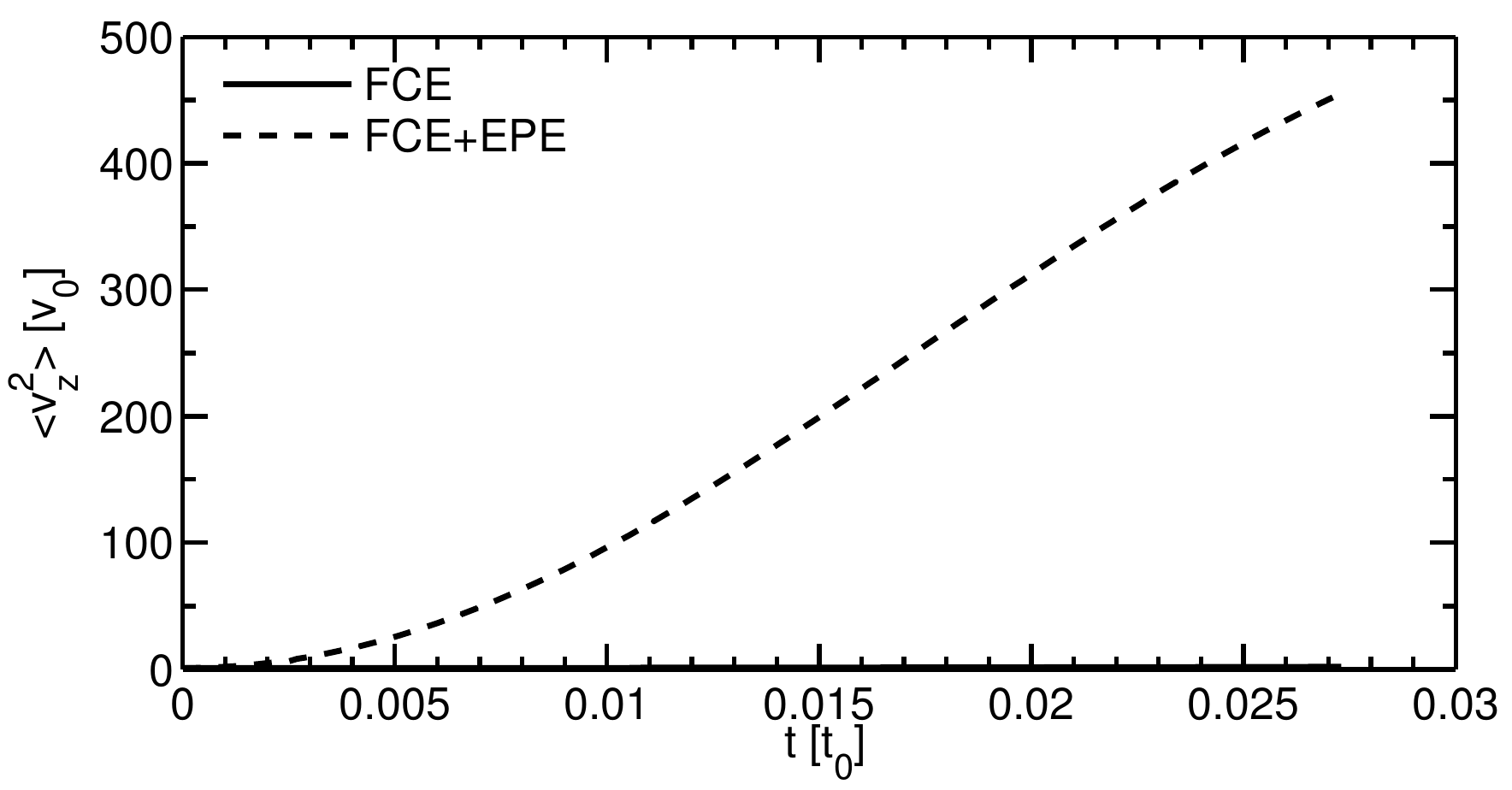}}
\caption{Electron mean square velocity as a function of time
  considering flow compressibility effects (FCE), and considering flow
  compressibility effects plus electron pressure effects (FCE+EPE): (Top)
Electron perpendicular velocity $v_\perp = \sqrt{v_x^2 + v_y^2}$ 
for FCE (solid line) and FCE+EPE (dashed line) for the case $M=0.25$. 
(Bottom) Mean square parallel ($v_z^2$) 
electron velocity. Note that
increase in FCE only case is small as in Fig 7}
\end{center}
\label{mean square velocity}
\end{figure}

In contrast a very different situation is observed for
electrons, as shown in Figure 9. The electron behavior is no longer
magnetized and the non constant perpendicular energy (top panel)
represents non-adiabatic motion. As seen in Fig. 9 (bottom panel) a
very high parallel energy (that is, high square parallel velocity) is 
reached in a short time, showing the importance of the EPE for
electrons in compressible MHD.

\section{\label{sec:level4}DISCUSSION:}
We investigated the effect of compressible MHD turbulence 
on particle energization, using test particle simulations in frozen
electromagnetic fields obtained from direct numerical solutions of the
MHD equations. We found that 
flow compressibility affects the energization of protons (i.e., in the
context of this work, test particles with gyroradius of the order of the 
MHD dissipation scale), while no significant effect is observed for
electrons (particles with gyroradius much smaller than the MHD
dissipation scale) as compared with the incompressible case.

Protons are accelerated by the perpendicular electric field generated on the interface of
current sheets, and they gain substantial energy as they encounter these structures. 
Moreover, the perpendicular electric field between current sheets is greater as compression
of the fluid increases, leading to a higher proton acceleration.

On the other hand, small gyroradii particles remain magnetized and
gain parallel energy as they travel along magnetic field lines almost
aligned with $B_0$. No effect of compressibility is noted for these
kind of particles and this is because the compressible modes in
magnetohydrodynamics are perpendicular propagating modes ($k \perp
B_0$). As a result no difference in the parallel electric field
obtained from static MHD fields is presented.

An interesting result is that when the model includes electron pressure gradients effects 
in the electric field, obtained from the generalized Ohm's law, one finds substantially greater parallel energization of electrons. In contrast, no 
significant changes are obtained for the proton energization with the inclusion
of the electron pressure gradient effects and of Hall currents.

The main aim of this paper was to analyze the case of weakly compressible turbulence, 
often appropriate to study the solar wind and other astrophysical scenarios, even 
though these plasmas can sometimes attain a strongly compressible state 
($M \geq 1$). We can thus conclude that at least for low turbulent 
Mach number, compression can enhance particle energization associated 
with coherent structures and therefore it has important implications for the study
of particle acceleration by turbulent fields. In the incompressible case, which is 
the limit of infinite sound wave velocity, protons can still be accelerated, but 
less than in the compressible case. The incompressible case thus served as a 
reference to measure the influence of compression on particle acceleration. Also, 
the incompressible case can still be relevant for some real physical
scenarios, such as the fast solar wind which might energize particles
as well\cite{Bogdan2014}.

We close with a remark concerning the importance of trapping effects in acceleration
of particles to higher energies in compressible turbulence. In general, for 
effective energization the particles must be exposed to a suitable electric field, 
but also the trajectory of the particle must allow  a long exposure time of the 
particle to the accelerating field. In the present case, parallel
acceleration of electrons occurs when their gyroradii 
are small compared to the width of mean field-aligned current
channels, as noted previously by \citet{PD1}. Analogous trapping
effects due to confinement in magnetic ``islands'' has been noted in
various systems from two dimensional MHD \cite{AmbrosianoEA88} to
fully kinetic PIC simulations\cite{DrakeEA06}. In those scenarios
small gyroradius particle are trapped for a period of time sufficient
for them to experience substantial parallel energization. Depending on
parameters this may be either heating (more particles, lower energies)
or acceleration (less particles but higher energy). On the other hand,
protons, having larger gyroradius, will not be easily trapped in
current channels, which often are a few proton inertial scales in width. 

The perpendicular acceleration mechanism described previously \cite{PD1,Dalena2012}
and elaborated on here, provides a way to accelerate protons (and heavier ions) due
to pependicular electric fields. The region of interaction between flux tubes 
provides the possibility of generating regions of effective
acceleration that may lie between reversing currents. Although 
these may be very complex regions in three dimensions, in a simplified two 
dimensional picture these can be flux pileup regions with gradients of the
perpendicular electric field. This transverse compression of the  magnetic field 
may occur even when the turbulence is incompressible. It is, however intuitively 
clear that compressibility will permit greater pileup and greater perpendicular
electric field gradients. In addition, to produce an efficient accelerator, the 
particles must also be trapped in the accelerating region for sufficient time. The
present numerical experiments also suggest that compressibility of the turbulence, 
acting near and within the regions between reversing currents, may provide 
substantially enhanced trapping for some particles. This is needed to explain the 
significantly greater perpendicular acceleration observed here when the turbulence
is compressible. 

\section*{Acknowldegments}

The authors thank an anonymous reviewer for useful comments that
resulted in the study of electron pressure effects. C.A.G, P.D., and
P.D.M. acknowledge support from grants UBACyT No. 20020110200359 and
20020100100315, and from grants PICT No. 2011-1529, 2011-1626, and
2011-0454. W.H.M. was partially supported by NASA LWS-TRT grant 
NNX15AB88G, Grand Challenge Research grant NNX14AI63G, and the Solar
Probe Plus mission through the Southwest Research Institute ISIS  
project D99031L.
\vspace*{1cm}
\nocite{*}
\bibliographystyle{unsrtnat}
\bibliography{aipsamp_arxiv}

\end{document}